\documentclass[aps,pra,twocolumn,groupedaddress,amsmath,amssymb]{revtex4-1}

\usepackage[usenames,dvipsnames]{xcolor}
\usepackage[colorlinks=true,citecolor=blue,linkcolor=blue,urlcolor=blue,backref=true]{hyperref}
\usepackage{amsmath}
\usepackage{amsfonts} 
\usepackage{amssymb}
\usepackage{latexsym}
\usepackage{mathrsfs}
\usepackage{bm}
\usepackage{nicefrac}
\usepackage{siunitx}
\usepackage{graphicx}
\usepackage{bm}
\usepackage{braket}
\usepackage{nicefrac}

\begin{document}

\title{The Casimir effect for fields with arbitrary spin}
\author{Adam Stokes and Robert Bennett}
\affiliation{The School of Physics and Astronomy, University of Leeds, Leeds, LS2 9JT, United Kingdom} 

\date{\today}

\begin{abstract}
The Casimir force between two perfectly reflecting parallel plates is considered. In a recent paper we presented generalised physical boundary conditions describing perfectly reflecting parallel plates. These boundary conditions are applicable to a field possessing any spin, and include the well-known spin-$\nicefrac{1}{2}$ and spin-$1$ boundary conditions as special cases. Here we use these general boundary conditions to show that the allowed values of energy-momentum turn out to be the same for any massless fermionic field and the same for any massless bosonic field. As a result one expects to obtain only two possible Casimir forces, one associated with fermions and the other with bosons. We explicitly verify that this is the case for the fields up to spin-$2$. A significant implication of our work is that periodic boundary conditions cannot be applied to a fermionic field confined between two parallel plates.
\end{abstract}

\maketitle

\section{Introduction}

In 1948 Casimir and Polder published a long, technically complex paper about the influence of retardation on the Van der Waals force \cite{Casimir:1948bd}. The stated goal of their work was to account for discrepancies between experiments and theory concerning colloidal suspensions of large particles \cite{verwey2001theory}. However, the work took on a whole new significance when, after discussing the results with Bohr, Casimir was inspired to try and re-explain his and Polder's results using the relatively new idea that the quantized electromagnetic field undergoes vacuum fluctuations. A short time later, Casimir published his now-famous paper \cite{casimir_attraction_1948} on the force of attraction between two infinite perfectly conducting parallel plates, whose presence modifies the quantized electromagnetic vacuum field. This force came to be known as the Casimir force. The calculation in \cite{casimir_attraction_1948} reproduces the results of  the much more involved calculation in \cite{Casimir:1948bd}, but is remarkable in its simplicity and elegance, while also providing one of the very few macroscopic manifestations of quantum field theory. The Casimir force is extremely weak so was initially nothing more than a theoretical curiosity, but as experimental methods improved the effect became measurable, and this led to rapidly increasing attention from the 1970s onwards. Since then there has been a profusion of extensions of Casimir's original work, which look into imperfectly conducting plates and different physical geometries \cite{LandauLifshitzPitaevskii, Boyer:1968ga, Rahi:2009fx, Schwinger:1978be,Philbin:2011di}. There have also been a number of experimental confirmations of the effect \cite{Lamoreaux:1997cu, Mohideen:1998ip,DeccaYukawa}. The existence of the Casimir effect is often cited in standard quantum field theory textbooks as the primary justification for the reality of vacuum fluctuations, though such interpretations carry some controversy \cite{Jaffe:2005km}. 

A fluctuating vacuum is a general feature of quantum fields, of which the free Maxwell field considered in \cite{Casimir:1948bd,casimir_attraction_1948,verwey2001theory, LandauLifshitzPitaevskii, Boyer:1968ga, Rahi:2009fx, Schwinger:1978be,Philbin:2011di, Lamoreaux:1997cu, Mohideen:1998ip,DeccaYukawa, Jaffe:2005km} is but one example. Fermionic fields such as that describing the electron, also undergo vacuum fluctuations, consequently one expects to find Casimir effects associated with such fields whenever they are confined in some way. Such effects were first investigated in the context of nuclear physics, within the so-called ``MIT bag model" of the nucleon \cite{chodos_new_1974}. In the bag-model one envisages the nucleon as a collection of fermionic fields describing confined quarks. These quarks are subject to a boundary condition at the surface of the `bag' that represents the nucleon's surface. Just as in the electromagnetic case, the bag boundary condition modifies the vacuum fluctuations of the field, which results in the appearance of a Casimir force \cite{Milton:1983ht, Milton:1983gg, Oxman:2005du, milonni_quantum_1994,fosco_functional_2008}. This force, although very weak at a macroscopic scale, can be significant on the small length scales encountered in nuclear physics. It therefore has important consequences for the physics of the bag-model nucleon \cite{ThomasWeiseStructNucleon}.  

The Maxwell and Dirac fields are both spinor fields, though the former is not usually described as such. It is possible to write Maxwell's equations in a form identical to the Dirac-Weyl equation that describes massless spin-$\nicefrac{1}{2}$ fermions \cite{BialynickiBirula:1996cf}. This naturally leads one to the question as to whether it is possible to use a spinor formalism to describe the Casimir effect for the Dirac (spin-$\nicefrac{1}{2}$) and the Maxwell (spin-$1$) fields in a unified way. We have shown \cite{stokes_unified_2014} that such a unification can be accomplished using the two-spinor calculus formalism introduced by Van der Waerden \cite{vanderWaerden:1928vw}. Moreover, this unification naturally lends itself to a generalization, which is applicable to confined higher-spin fields. These fields include the spin-$2$ field associated with the so-called graviton, which appears in linearized quantum gravity, and its supersymmetric partner the spin-$\nicefrac{3}{2}$ gravitino.

In this paper we will present specific results for the Casimir force associated with the fields up to spin-$2$. We organize our work by noting that calculations of Casimir forces broadly follow the following three steps: 

\begin{enumerate}
\item The statement of one or more boundary conditions governing how the considered field behaves at material surfaces. These can be mathematically convenient (examples include Dirichlet  \cite{Graham:2004wp}, Neumann \cite{Alves:2003jk}, Robin \cite{Romeo:2002ef} and periodic \cite{wen-biao_casimir_2007,Langfeld:1995cn} BCs) or physically-motivated (those imposed by electromagnetism \cite{LandauLifshitzPitaevskii} or by the bag model \cite{chodos_new_1974} for example). 
\item The determination of a set of field solutions that obey the boundary conditions specified in step 1. In the simplest cases this can be achieved by direct solution of the equations of motion. However, this step is usually non-trivial, and has resulted in the development of numerous techniques including the so-called macroscopic QED \cite{Dung:1998ic}, worldline numerics \cite{Gies:2006fp}, the ``proximity-force approximation" \cite{Derjaguin:1934ec,Gies:2006di}, certain scattering theory based methods \cite{Jaekel:1991bn}, and many more.
\item The substitution of the field solutions found in step 2 into an expression for the vacuum energy of the relevant field. Upon suitable regularisation and the dropping of any boundary-independent terms, one is left with the Casimir force for some combination of: a field (Maxwell, Dirac, etc), a boundary condition, and a physical geometry. 
\end{enumerate}

We will begin in section \ref{GBCs} by reviewing the generalized, physically-motivated boundary conditions presented in \cite{stokes_unified_2014}. We then find explicit field solutions for the parallel-plate geometry, and hence accomplish steps $1$ and $2$ given above. In section \ref{CasimirSection} we will carry out the final step above by computing specific values for the Casimir force associated with the massless fields up to spin-$2$.

\section{Generalised physical boundary conditions}\label{GBCs}

In this section we review our generalisation of the boundary conditions (BCs) employed in the calculation of the Casimir effect associated with the spin-$\nicefrac{1}{2}$ and spin-$1$ fields. To do this we use the two-spinor calculus formalism presented in appendix \ref{A}. Further details of the two-spinor calculus formalism can be found, for example, in \cite{penrose_spinors_1987,barut_electrodynamics_1964,dreiner_two-component_2010} and the references therein.

\subsection{Unified physical boundary conditions for massless spin-$\nicefrac{1}{2}$ and spin-$1$ fields}

\subsubsection{spin-$\nicefrac{1}{2}$}

We begin by considering the simplest spin field---the spin-$\nicefrac{1}{2}$ massless Dirac-Weyl field. We adopt the two-spinor calculus formalism laid out in appendix \ref{A}. The spin-$\nicefrac{1}{2}$ massless field is described by a pair of square-root Klein-Gordon equations, which in the massless case are decoupled;
\begin{align}\label{eqnmohalf}
{\sigma^\mu}_{{\bar a}a}\partial_\mu \psi^a =0,~~~~~~{\tilde \sigma}^{\mu a{\bar a}}\partial_\mu \phi_{\bar a} =0
\end{align}
where $\psi^a$ and $\phi_{\bar a}$ are spin-$\nicefrac{1}{2}$ massless quantum fields describing right and left-helicities respectively. The usual Dirac bispinor can be constructed through a direct sum of fields proportional to $\psi$ and $\phi$.

\begin{figure}[h]
\center
\includegraphics[width=0.8\columnwidth]{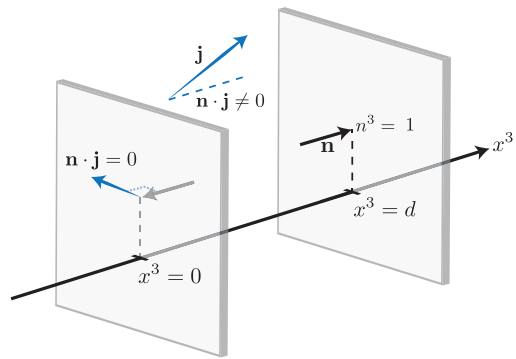} 
\caption{Infinite parallel plates located at $x^3 = 0$ and $x^3 =d>0$. The outward pointing normals to the plates are shown. Note that since the two normals point in opposite directions their components differ by a minus sign at each plate; $-n^3(0)=n^3(d)=1=n_3(0)=-n_3(d)$. The physical constraint that at the plates there is no current normal to the surface is illustrated.}\label{plates}
\end{figure}
We wish to calculate the Casimir effect associated with the massless spin-$\nicefrac{1}{2}$ field due to the presence of two perfectly reflecting parallel plates orthogonal to the $x^3$-axis. We assume that one plate is located at $x^3=0$ and that the other is located at $x^3=d>0$ as shown in Fig. \ref{plates}. The physical constraint we impose on the fields is that there be no particle-current normal to the surfaces at $x^3=0,d$ \cite{milonni_quantum_1994};
\begin{align}\label{cascurrent}
n_\mu(0,d) j^\mu(0,d) =0
\end{align}
where $n^\mu(0,d)=(0,0,0,\mp 1)$ are components of the outward-pointing unit normals to the surfaces at $0$ and $d$ respectively, and $j^\mu$ are the components of the spin-$\nicefrac{1}{2}$ particle-current vector. In terms of the two-spinors in Eq. (\ref{eqnmohalf}), Eq. (\ref{cascurrent}) reads [c.f. Eq. (\ref{sv})]
\begin{align}\label{cascurrenthalf}
n_{{\bar a}a}(0,d)j^{a{\bar a}}(0,d) = n_{{\bar a}a}(0,d)\psi^a(0,d)\psi^{\bar a}(0,d) = 0.
\end{align}
This condition will hold if we impose the BCs \cite{milonni_quantum_1994}
\begin{align}\label{bchalf}
n_\mu(0,d){\sigma^\mu}_{{\bar a}a}(0,d)\psi^a(0,d)\equiv \pm{\sigma^3}_{{\bar a}a}\psi^a(0,d)=\psi_{\bar a}(0,d),
\end{align}
where the (normalised) Pauli matrices $\{\sigma^\mu\}$ are defined in Eq. (\ref{Pauli}). The $+$ sign in Eq. (\ref{bchalf}) corresponds to the case $x^3=0$, and the $-$ sign to the case $x^3=d$. To prove that Eq. (\ref{bchalf}) implies Eq. (\ref{cascurrenthalf}), one first multiplies Eq. (\ref{bchalf}) through by $\psi^{\bar a}(0,d)$. The left-hand-side then equals $n_\mu(0,d)j^\mu(0,d)$ while the right-hand-side equals $\psi_{\bar a}(0,d)\psi^{\bar a}(0,d)={\bar \omega}(\psi(0,d),\psi(0,d))\equiv 0$, where ${\bar \omega}$ is the symplectic form on the spinor-space ${\bar S}$ [c.f. appendix \ref{A}]. This completes the proof. We note that the BCs in Eq. (\ref{bchalf}) are nothing but the usually employed BCs in the calculation of the spin-$\nicefrac{1}{2}$ Casimir effect \cite{chodos_new_1974,milonni_quantum_1994}.

We now Fourier-expand the fields $\psi^a$ and $\phi_{\bar a}$ in plane-wave superpositions as follows
\begin{align}\label{fexhalf}
&\psi^a(x) = \int d^3k \, u^a({\bf k})\hspace*{-0.3mm}\left(b_R({\bf k})e^{-ik_\mu x^\mu}+d^\dagger_{L}({\bf k})e^{ik_\mu x^\mu}\right)\nonumber \\
&\phi_{\bar a}(x) =\int d^3k \, u_{\bar a}({\bf k})\hspace*{-0.3mm}\left(d_R({\bf k})e^{-ik_\mu x^\mu}+b^\dagger_L({\bf k})e^{ik_\mu x^\mu}\right).
\end{align}
Here $k^0 \equiv \omega := |{\bf k}|$, and the label $R$ (for right) corresponds to the helicity $\lambda =\nicefrac{1}{2}$ while the label $L$ corresponds to the helicity $\lambda=-\nicefrac{1}{2}$. The function $u^a({\bf k})$ and its charge-conjugate $u_{\bar a}({\bf k})$ are (suitably normalised) momentum-space single-particle right and left-helicity wavefunctions respectively. The $b_\lambda({\bf k})$ and $b^\dagger_\lambda({\bf k})$ are annihilation and creation operators for particles with momentum ${\bf k}$ and helicity $\lambda$, and the $d_\lambda({\bf k})$ and $d^\dagger_\lambda({\bf k})$ are the corresponding anti-particle operators. Altogether these operators satisfy the fermionic anti-commutation relations
\begin{align}\label{acom}
\{b_\lambda({\bf k}),b^\dagger_{\lambda'}({\bf k}')\} = \{d_\lambda({\bf k}),d^\dagger_{\lambda'}({\bf k}')\} =\delta_{\lambda \lambda'}\delta({\bf k}-{\bf k}').
\end{align}
The expansions in Eq. (\ref{fexhalf}) and anti-commutation relations in Eq. (\ref{acom}) allow us to express the total energy of the spin-$\nicefrac{1}{2}$ massless field (including both helicities) as
\begin{align}\label{Hhalf}
H_{\nicefrac{1}{2}}=\int d^3k \sum_{\lambda=\pm \nicefrac{1}{2}} \omega\left(b^\dagger_\lambda({\bf k})b_\lambda({\bf k}) + d^\dagger_\lambda({\bf k})d_\lambda({\bf k}) -1\right).
\end{align}

In order to deduce the values of momentum ${\bf k}$ allowed by the BCs in Eq. (\ref{bchalf}) it suffices to consider the following positive-energy single-particle plane-wave solution of the Dirac-Weyl equation (\ref{eqnmohalf}) for $\psi^a$;
\begin{align}\label{fr}
\psi^a(x)=e^{-i\omega t}\psi^a({\bf x}),~~~~ \psi^a({\bf x}) = u^a({\bf k})e^{i{\bf k}\cdot {\bf x}}.
\end{align}
If this solution is to be non-trivial, i.e., such that $u^a({\bf k})$ is not identically zero, then it cannot satisfy the BCs in Eq. (\ref{bchalf}). In other words the possibility of a completely free solution is negated by the presence of the plates, which evidently must modify the free solution in some way. We make the ansatz \cite{chodos_new_1974,milonni_quantum_1994}
\begin{align}\label{msoln}
\psi^a({\bf x}) = \left(\sigma^{0 a{\bar a}} e^{i{\bf k}_\perp \cdot {\bf x}_\perp} + \sigma^{3 a{\bar a}} e^{-ik^3x^3}\right)u_{\bar a}({\bf k})
\end{align}
where ${\bf v}_\perp :=(v^1,v^2,0)$ denotes the projection of the three-vector ${\bf v}$ onto the $x^1$-$x^2$ plane. Making this ansatz constitutes the physically reasonable assumption that the effect of the plates is to flip the sign of $x^3$ in Eq. (\ref{fr}). The modified solution (\ref{msoln}) satisfies the BC in Eq. (\ref{bchalf}) at $x^3=0$ identically. It will also satisfy the BC corresponding to $x^3=d$ provided that
\begin{align}
\cos(k^3 d) = 0.
\end{align}
Thus, given the solution in Eq. (\ref{msoln}) the BCs in Eq. (\ref{bchalf}) restrict the values of $k^3$ and hence ${\bf k}$ to
\begin{align}\label{khalf}
k^3 = {n\pi \over 2d},~~~~ {\bf k} = \left(k^1,k^2,{n\pi\over 2d}\right),~~~~n=1,3,5,...\, .
\end{align}
When used in conjunction with Eqs. (\ref{Hhalf}) and (\ref{khalf}), the above restricted values of ${\bf k}$ yield the usual Casimir force associated with the spin-$\nicefrac{1}{2}$ field between two perfectly reflecting parallel plates (c.f. section \ref{cas1/2}). In section \ref{genbc} it will be shown that with a straightforward generalisation of the BCs in Eq. (\ref{bchalf}), the above values of ${\bf k}$ turn out to be the same for any fermionic field. 

\subsubsection{Spin-$1$}\label{1bcsec}

The massless spin-$1$ field is the familiar field of Maxwell electrodynamics. In the two-spinor calculus formalism the Maxwell field is described by a pair of symmetric spin-tensors $\psi^{ab}$ and $\psi_{{\bar a}{\bar b}}$ corresponding to the right and left-helicity states of the photon respectively. The equations of motion analogous to those in (\ref{eqnmohalf}) are the square-root Klein-Gordon equations
\begin{align}\label{me}
{\sigma^\mu}_{{\bar a}a}\partial_\mu \psi^{ab} = 0,~~~~~~{\tilde \sigma}^{\mu a{\bar a}}\partial_\mu \psi_{{\bar a}{\bar b}}=0.
\end{align}

The relation of the spin-tensors above to the more conventional electromagnetic three-vectors is most easily achieved through the complex Riemann-Silberstein vector ${\bf F}:={\bf E}+i{\bf B}$, where ${\bf E}$ and ${\bf B}$ are the electric and magnetic fields respectively. The vectors ${\bf F}$ and ${\bf F}^*$ can be viewed as complex three-vectors corresponding to the right and left-helicity states of the photon \cite{bialynicki-birula_role_2013}. If one replaces the imaginary unit $i$ in these definitions with the volume form $I$ on Minkowski space-time $E^{1,3}$ and views the fields ${\bf E}$ and ${\bf B}$ as bivectors $E$ and $B$ over $E^{1,3}$, then one obtains the familiar electromagnetic field tensor $F\equiv E+IB$ and its reverse $F^\dagger \equiv E-IB$. Here the juxtaposition $IB$ denotes the Clifford (geometric) product of $I$ and $B$ \cite{doran_geometric_2003}. One can also describe the electromagnetic field in terms of the dual tensor $G:=\ast F\equiv -IF$ where $\ast$ denotes the Hodge-dual. Clearly each of the objects $\psi^{ab}$, ${\bf F}$ and $F$ (or $G$) constitutes a different organisation of the six real degrees of freedom (including gauge degrees of freedom) that describe the physical electromagnetic field.

Relating the Maxwell spin-tensor $\psi^{ab}$ to the Riemann-Silberstein vector ${\bf F}$ we have \cite{bialynicki-birula_role_2013}
\begin{align}\label{rstomspin}
\psi^{ab} = \left( {\begin{array}{cc}
-F^1+iF^2& F^3 \\
 F^3 & F^1+iF^2  \\
 \end{array} } \right),
\end{align}
and relating $\psi^{ab}$ to the electromagnetic bivector $F^{\mu\nu}$ we have [c.f. Eq. (\ref{sv})] \cite{penrose_spinors_1987}
\begin{align}
F^{\mu\nu}&={\sigma^\mu}_{{\bar a}a}{\sigma^\nu}_{{\bar b}b}F^{a{\bar a}b{\bar b}}, & F^{a{\bar a}b{\bar b}}&=\omega^{{\bar a}{\bar b}}\psi^{ab} + \omega^{ab}\psi^{{\bar a}{\bar b}}.
\end{align}
Although either of the above relations suffices in order to relate the two-spinor treatment of electrodynamics to the more widely-known approaches, we choose to make use of the former relation (\ref{rstomspin}). Using Eq. (\ref{rstomspin}) it is straightforward to show that the equations in (\ref{me}) are equivalent to the free Maxwell equation
\begin{align}
i\partial_t{\bf F}=\nabla\times {\bf F}
\end{align}
and its complex-conjugate.

In determining the appropriate BCs to impose for the calculation of the Casimir effect in the spin-$1$ case, one must contend with the fact that there exists no local particle-current vector for massless fields with spin greater than $\nicefrac{1}{2}$ \cite{penrose_zero_1965}. Thus, for the Maxwell field in particular, an alternative physical current $j^\mu$ must be chosen in order to obtain a condition analogous to Eq. (\ref{cascurrent}). As is well-known, one of the few local observables associated with photons is their energy-density \cite{sipe_photon_1995,bialynicki-birula_role_2013}. A natural choice for $j^\mu$ in the spin-$1$ case is therefore the energy-current
\begin{align}\label{onecurrent}
j^\mu := T^{\mu 0} &= {1\over 2}\left({\bf F}\cdot{\bf F}^*, i{\bf F}\times{\bf F}^*\right) \nonumber \\ &=\left({1\over 2}\left({\bf E}^2+{\bf B}^2\right),{\bf E}\times{\bf B}\right)
\end{align}
where $T^{\mu\nu}$ are the components of the symmetric energy-momentum tensor of the Maxwell field. The physical constraint expressed by Eq. (\ref{cascurrent}) with $j^\mu$ given by Eq. (\ref{onecurrent}) is ensured if one assumes that
\begin{align}\label{cas1}
{\bf n}(0,d)\cdot {\bf B}(0,d) =&\,0,& {\bf n}(0,d)\times {\bf E}(0,d)=&\,0
\end{align}
where as before ${\bf n}(0)$ and ${\bf n}(d)$ are normal to the surfaces at $x^3=0$ and $d$ respectively. The analogy between the spin-$\nicefrac{1}{2}$ and spin-$1$ cases becomes obvious when Eqs. (\ref{onecurrent}) and (\ref{cas1}) are written in terms of the Maxwell spin-tensor $\psi^{ab}$. To this end we note that using Eq. (\ref{rstomspin}) the energy-momentum tensor $T^{\mu \nu}$ can be written [c.f. Eq. \ref{sv}]
\begin{align}\label{mspincurrent}
T^{\mu\nu} = {\sigma^\mu}_{{\bar a}a}{\sigma^\nu}_{{\bar b}b}T^{a{\bar a}b{\bar b}},~~~~T^{a{\bar a}b{\bar b}} = \psi^{ab}\psi^{{\bar a}{\bar b}}.
\end{align}
Within the two-spinor calculus formalism there is a clear analogy between the spin-$1$ energy-momentum spin-tensor $T^{a{\bar a}b{\bar b}} =\psi^{ab}\psi^{{\bar a}{\bar b}}$, and the spin-$\nicefrac{1}{2}$ particle-current $j^{a{\bar a}}=\psi^a\psi^{\bar a}$. As such, substituting the expression for $T^{\mu 0}$ given by Eq. (\ref{mspincurrent}) with $\nu\equiv 0$, into Eq. (\ref{cascurrent}) yields
\begin{align}
&n_\mu(0,d) j^\mu (0,d) =n_{{\bar a}a}(0,d){\sigma^0}_{{\bar b}b}T^{a{\bar a}b{\bar b}}(0,d) \nonumber \\ & \equiv n_{{\bar a}a}(0,d){\sigma^0}_{{\bar b}b}\psi^{ab}(0,d)\psi^{{\bar a}{\bar b}}(0,d) =0,
\end{align}
which is the spin-$1$ version of Eq. (\ref{cascurrenthalf}). This condition will necessarily hold if we impose the BCs
\begin{align}\label{bcone}
n_\mu(0,d)n_\nu(0,d) {\sigma^\mu}_{{\bar a}a} {\sigma^\nu}_{{\bar b}b} \psi^{ab}(0,d) &\equiv {\sigma^3}_{{\bar a}a} {\sigma^3}_{{\bar b}b}\psi^{ab}(0,d) \nonumber \\ &=\psi_{{\bar a}{\bar b}}(0,d) 
\end{align}
which is clearly the spin-$1$ version of Eq. (\ref{bchalf}). To see that Eqs. (\ref{cas1}) and (\ref{bcone}) are equivalent one need only expand the sums in Eq. (\ref{bcone}), which gives
\begin{align}\label{mbcex}
\psi^{00}(0,d)=\psi_{{\bar 0}{\bar 0}}(0,d),~~~~~~\psi^{01}(0,d)=-\psi_{{\bar 0}{\bar 1}}(0,d).
\end{align}
Using Eq. (\ref{rstomspin}) it is easy to show that these conditions are equivalent to Eq. (\ref{cas1}). The BCs in Eq. (\ref{bcone}) written in terms of $\psi^{ab}$, are therefore completely equivalent to the usual BCs [in Eq. (\ref{cas1})] employed in the calculation of the electromagnetic Casimir effect \cite{casimir_attraction_1948}.

There still remains the proof that the BCs in Eq. (\ref{bcone}) imply $n_\mu(0,d)T^{\mu0}(0,d)=0$. This proof is not as straightforward as the proof given in the spin-$\nicefrac{1}{2}$ case. There we were able to make use of the anti-symmetry of the symplectic form on the spinor space ${\bar S}$, but such a strategy will obviously fail if the field under consideration is bosonic, i.e., is described by an evenly ranked spin-tensor. However, a more brute force proof involving the use of Eq. (\ref{mbcex}) is available in the spin-$1$ case. In section \ref{genbc} we specify generalised BCs associated with an arbitrary spin-$\nicefrac{m}{2}$ field [Eq. (\ref{gen})], and prove in appendix \ref{B} that they imply the vanishing of the normal component of the relevant local current [Eq. (\ref{currentm})]. In this context the Maxwell field simply corresponds to the special case $m=2$.

Before moving on to determine the allowed values of ${\bf k}$ for the spin-$1$ field, we note that a significant difference between Eq. (\ref{bcone}) and Eq. (\ref{bchalf}), is that Eq. (\ref{bcone}) involves an even number of factors of $n_\mu$ rather than an odd number. Thus, the boundary condition is the same for both the $x^3=0$ and $x^3=d$ cases, rather than differing by a minus sign as is the case in Eq. (\ref{bchalf}) for the spin-$\nicefrac{1}{2}$ field. We will see quite generally in section \ref{genbc}, that this is the crucial difference between the BCs for fermionic (half odd-integer spin) and bosonic (integer-spin) fields.

Now, using Eq. (\ref{bcone}) the determination of the allowed values of $k^3$ in the spin-$1$ case exactly mirrors the procedure used above for the spin-$\nicefrac{1}{2}$ field. We employ the usual expression for the quantised energy of the Maxwell field
\begin{align}\label{Hone}
H_1:=T^{00} = \int d^3k \sum_{\lambda =\pm 1} \omega \left(a^\dagger_\lambda({\bf k})a_\lambda({\bf k}) + {1\over 2}\right)
\end{align}
where $a_\lambda({\bf k})$ and $a^\dagger_\lambda({\bf k})$ are bosonic annihilation and creation operators satisfying the commutation relation
\begin{align}\label{com}
[a_\lambda({\bf k}),a^\dagger_{\lambda'}({\bf k}')]=\delta_{\lambda\lambda'}\delta({\bf k}-{\bf k}').
\end{align}
In order to determine the values of ${\bf k}$ allowed by the BCs in Eq. (\ref{bcone}), we consider in exact analogy to the spin-$\nicefrac{1}{2}$ case, the following modified positive-energy solutions of the Maxwell equations (\ref{me})
\begin{align}
&\psi^{ab}(x) = e^{-i\omega t}\psi^{ab}({\bf x}), \nonumber \\
&\psi^{ab}({\bf x}) = \left(\sigma^{0 a{\bar a}} \sigma^{0 b{\bar b}}e^{i{\bf k}_\perp \cdot {\bf x}_\perp} + \sigma^{3 a{\bar a}} \sigma^{3 b{\bar b}}e^{-ik^3x^3}\right)u_{{\bar a}{\bar b}}({\bf k})
\end{align}
where $u^{ab}({\bf k})$ is a momentum-space left-helicity single-particle spin-tensor, analogous to $u^a({\bf k})$ for the spin-$\nicefrac{1}{2}$ field. As in the spin-$\nicefrac{1}{2}$ case the modified solution satisfies the BC in Eq. (\ref{bcone}) at $x^3=0$ identically. It will also satisfy the BC at $x^3=d$ provided
\begin{align}
\sin(k^3 d)=0,
\end{align}
which is the case if and only if
\begin{align}\label{kone}
k^3={n\pi \over d},~~~~{\bf k} = \left(k^1,k^2,{n\pi \over d}\right),~~~~n=0,1,2,...\, .
\end{align}
These values of ${\bf k}$ are usually found by more conventional means involving the electric and magnetic fields and the BCs in Eq. (\ref{cas1}). Along with Eq. (\ref{Hone}) they can be used to obtain the well-known electromagnetic Casimir force first found in \cite{casimir_attraction_1948} (c.f. section \ref{cas1sec}). In the following section (\ref{genbc}) we generalise the BCs in Eq. (\ref{bcone}) to show that the above values of ${\bf k}$ turn out to be the allowed values for any bosonic field.

\subsection{Generalised physical boundary conditions for massless fields with higher spin}\label{genbc}

\begin{figure}[t]
\center
\includegraphics[width=\columnwidth]{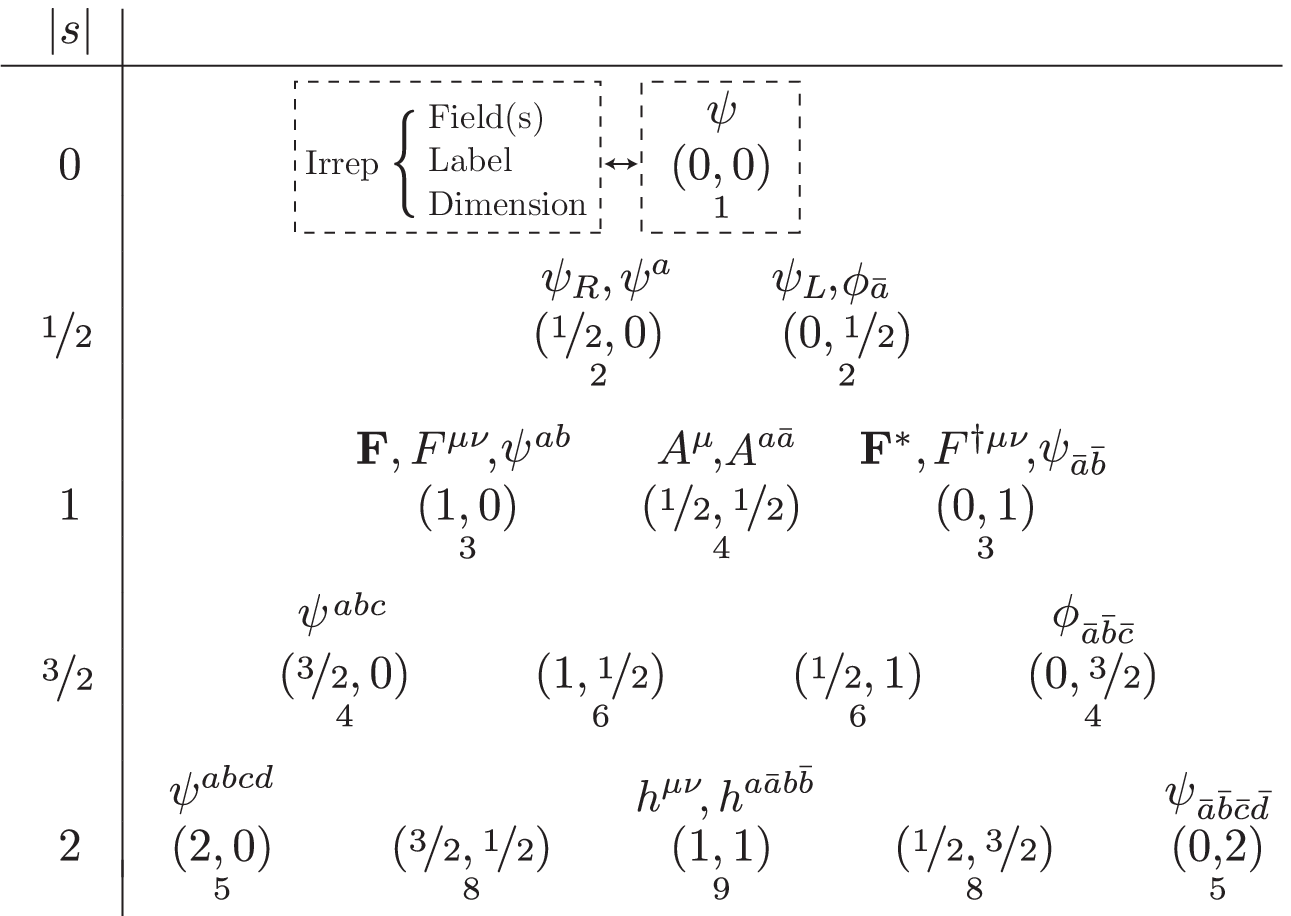}
\caption{A diagrammatic layout of the irreducible representations (irreps) of $Sp(2,{\mathbb C})$ for massless fields with discrete spin. Each irrep is labelled by a pair of numbers $(i,j)$ where $i,j=0,\nicefrac{1}{2},1,\nicefrac{3}{2},...\,$. The spin values corresponding to the $(i,j)$'th-representation are $\pm |s|$ where $|s|=i+j$. An $(i,j)$-irrep for which $i=0$ or $j=0$ is called an ``outer" irrep, otherwise it is an ``inner" irrep. Fields belonging to the outer irrep $(i,0)$ directly describe a spin $i$ particle, whereas those belonging to the outer irrep $(0,i)$ directly describe a spin $-i$ particle. The number directly below each $(i,j)$-irrep equals the complex-dimension $d=(2i+1)(2j+1)$ of the irrep immediately above it. Directly above the relevant $(i,j)$-irreps is a list of fields. Under a Lorentz transformation, each of these fields transforms in some way according to the irrep below it (c.f. appendix \ref{Lts}). For example, in the two-spinor calculus formalism a field belonging to the $(i,j)$'th-irrep is denoted ${\psi^{(a_1...a_{2i})}}_{({\bar a}_1...{\bar a}_{2j})}$ where $(a_1...a_n)$ denotes total symmetrisation of the indices $a_1,...,a_n$ (c.f. appendix \ref{Lts}). Since lower spins have been studied more extensively than higher spins, lower spin values possess more corresponding fields. This is particularly evident in the electromagnetic case $|s|=1$ (c.f. section \ref{1bcsec}). For $s=0$ the only corresponding field is the scalar field $\psi$. For $|s|=\nicefrac{1}{2}$ we have in addition to $\psi^a$ and $\phi_{\bar a}$, the Dirac-Weyl two-column spinors $\psi_R$ and ${\psi_L}$, which make up the usual Dirac bispinor. The irreps of the form $(i,i)$ correspond to totally symmetric traceless rank $2i \choose 2i$-tensors over Minkowski spacetime (c.f. appendix \ref{A3}). Such tensors represent field potentials. The first of the $(i,i)$ irreps is the $(\nicefrac{1}{2},\nicefrac{1}{2})$ vector irrep occuring for $|s|=1$. It has corresponding field potential $A^\mu$, which denotes the electromagnetic four-potential. The next is the $(1,1)$ irrep occuring for $|s|=2$, and corresponding to which $h^{\mu\nu}$ is the field potential describing linearised gravity (c.f. section \ref{hspins2}).} \label{figreps}
\end{figure}

Given the analogy between the spin-$\nicefrac{1}{2}$ and spin-$1$ fields described above, the extension of the physical BCs in Eqs. (\ref{bchalf}) and (\ref{bcone}) to higher-spin fields now naturally presents itself. Initially for higher spin fields the identification of a local physical current to use in conjunction with Eq. (\ref{cascurrent}) may seem problematic, but we have effectively already tackled this problem in adapting the spin-$\nicefrac{1}{2}$ BCs to the spin-$1$ case. The BCs for any massless higher spins should evidently involve the physical fields directly involved in the description of the right and left-helicity states. These fields belong to the (carrier spaces of the) ``outer" (rather than the ``inner") irreducible representations of the symplectic group $Sp(2,{\mathbb C})$. This terminology is clarified by Fig. \ref{figreps}. The physical fields belonging to the outer representations are those that appear explicitly in the square-root (massless) Klein-Gordon equations, which describe the physical dynamics of the free system. For a spin-$\nicefrac{m}{2}$ field these equations read
\begin{align}\label{geneqnmo}
{\sigma^\mu}_{{\bar a}_1a_1}\partial_\mu \psi^{a_1 a_2 a_3 ... a_m} = 0,~~~~~{\tilde \sigma}^{\mu a_1{\bar a}_1}\partial_\mu \phi_{{\bar a}_1 {\bar a}_2 {\bar a}_3 ... {\bar a}_m} = 0.
\end{align}
Specifying the appropriate BCs in terms of the outer field $\psi^{a_1...a_m}$ clearly avoids any discussion regarding the use of unphysical inner field potentials, which are often used in the description of higher spins.

Generalising the local currents encountered in the spin-$\nicefrac{1}{2}$ and spin-$1$ cases, we begin by defining for a spin-$\nicefrac{m}{2}$ field the local current
\begin{align}\label{currentm}
j^\mu(m) := {\sigma^\mu}_{{\bar a}_1a_1} {\sigma^0}_{{\bar a}_2a_2}... {\sigma^0}_{{\bar a}_ma_m}\psi^{a_1...a_m}\psi^{{\bar a}_1...{\bar a}_m},
\end{align}
which due to the first equation of motion in (\ref{geneqnmo}), satisfies the local continuity equation $\partial_\mu j^\mu(m)=0$. Generalising now the spin-$\nicefrac{1}{2}$ and spin-$1$ BCs in Eqs. (\ref{bchalf}) and (\ref{bcone}), we impose for a spin-$\nicefrac{m}{2}$ field the BCs
\begin{align}\label{gen}
&n_{\mu_1}(0,d)...n_{\mu_m}(0,d){\sigma^{\mu_1}}_{{\bar a}_1 a_1}...{\sigma^{\mu_m}}_{{\bar a}_m a_m}\psi^{a_1 ... a_m}(0,d) \nonumber \\&= \psi_{{\bar a}_1...{\bar a}_m}(0,d),
\end{align}
which are the same generalised BCs first given in \cite{stokes_unified_2014}. We prove in appendix \ref{B} that these BCs imply that $j^\mu(m)$ satisfies the physical constraint given in Eq. (\ref{cascurrent}), i.e., $n_\mu(0,d)j^\mu(0,d;m)=0$. For fermionic fields $m$ is odd, which implies that the BCs in Eq. (\ref{gen}) contain an odd number of factors of $n_\mu(0,d) =(0,0,0,\pm 1)$. This means that when $x^3=d$ the BC differs by a minus sign compared with when $x^3=0$. In contrast, for bosonic fields $m$ is even, so the BC is the same for both the $x^3=d$ and $x^3=0$ cases.

\begin{figure}[t]
\center
\includegraphics[width=\columnwidth]{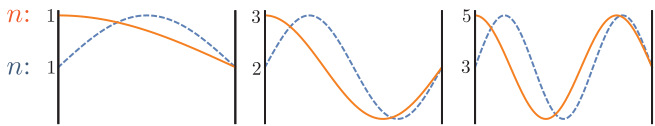} 
\caption{A schematic representation of the field solutions corresponding to the first few allowed values of $k^3$ from Eq. (\ref{ks}). Fermionic field solutions corresponding to the three values $k^3 =n\pi/2d,~n=1,3,5$ are represented by orange curves, while bosonic field solutions corresponding to the three values $k^3 =n\pi/d,~n=1,2,3$ are represented by dashed blue curves. The aperiodic orange curves clearly illustrate that periodic BCs cannot be used in conjunction with fermionic fields.}\label{SetupFig}
\end{figure}

To determine the allowed values of $k^3$ due to the BCs in Eq. (\ref{gen}), we use in analogy to the spin-$\nicefrac{1}{2}$ and spin-$1$ cases, the following modified single-particle positive-energy solution to the first of the equations in (\ref{geneqnmo})
\begin{align}
&\psi^{a_1...a_m}(x)=e^{-i\omega t}\psi^{a_1...a_m}({\bf x}),\nonumber \\
&\psi^{a_1...a_m}({\bf x}) := \Big( \sigma^{0 a_1{\bar a}_1}...\sigma^{0a_m{\bar a}_m}e^{i{\bf k}_\perp \cdot {\bf x}_\perp}  \nonumber \\ & \hspace*{2.2cm}+ \sigma^{3 a_1{\bar a}_1}...\sigma^{3a_m{\bar a}_m}e^{-ik^3 x^3}\Big) u_{{\bar a}_1...{\bar a}_m}({\bf k})
\end{align}
where $u^{a_1...a_m}({\bf k})$ is a momentum-space single-particle positive-energy spin-tensor. In complete analogy to the spin-$\nicefrac{1}{2}$ and spin-$1$ cases, substituting the solution above into Eq. (\ref{gen}) implies
\begin{align}
&\cos(k^3 d)=0,~~~~~m~{\rm odd}\nonumber \\
&\sin(k^3 d)=0,~~~~~\,m~{\rm even}.
\end{align}
This gives the allowed values of $k^3$ and ${\bf k}$ as
\begin{align}\label{ks}
&k^3 = {n\pi \over 2d},~~~~{\bf k} = \left(k^1,k^2,{n\pi\over 2d}\right),~~n=1,3,5,...~~m~{\rm odd}\nonumber \\
&k^3={n\pi \over d},~~~~{\bf k} = \left(k^1,k^2,{n\pi \over d}\right),~~n=0,1,2,...~~m~{\rm even}.
\end{align}
In Fig. \ref{SetupFig} we give a schematic representation of field solutions corresponding to the first few allowed values of $k^3$ specified above.

According to the generalised BCs in Eq. (\ref{gen}) the two sets of energy-momentum values in (\ref{ks}) are the only two possible, and they correspond to fermionic and bosonic fields respectively. A significant implication of these general results is that periodic BCs cannot be applied to a physically confined fermionic field between two parallel plates, because for fermionic fields the BCs are necessarily different at the two surfaces as is lucidly illustrated in Fig. \ref{SetupFig}. This means for example, that the approach adopted in \cite{wen-biao_casimir_2007} where periodic BCs were imposed on the spin-$\nicefrac{3}{2}$ field, is unphysical.

\section{The Casimir effect for arbitrary spin fields}\label{CasimirSection}

Having obtained according to the BCs in Eq. (\ref{gen}), the allowed values of energy-momentum for a field with arbitrary spin, we now look to calculate the resulting Casimir force. 

\subsection{The Casimir effect for spin-$\nicefrac{1}{2}$ and spin-$1$ fields}

Calculating the Casimir forces associated with the spin-$\nicefrac{1}{2}$ and spin-$1$ fields is a textbook exercise, so we review it here only very briefly, with an eye towards extending the calculation to the case of higher spin fields.

\subsubsection{spin-$\nicefrac{1}{2}$}\label{cas1/2}

The energy associated with the massless spin-$\nicefrac{1}{2}$ field is given in Eq. (\ref{Hhalf}). Our strategy in calculating the Casimir force is to initially restrict the field to a fictitious cavity upon which we impose periodic BCs. We will take the continuum limit in the $k^1$ and $k^2$-directions after having employed the allowed values of ${\bf k}$ given in Eq. (\ref{khalf}). The vacuum energy inside the quantisation cavity is given by
\begin{align}\label{evachalf}
E_{\nicefrac{1}{2}}^{\rm vac} = -\sum_{\bf k}\sum_{\lambda=\pm\nicefrac{1}{2}} \omega
\end{align}
where $\omega = |{\bf k}|$. Substituting into Eq. (\ref{evachalf}) the allowed values of ${\bf k}$ given in Eq. (\ref{khalf}) and taking the continuum limit with respect to the $k^1$ and $k^2$-directions one obtains \cite{milonni_quantum_1994}
\begin{align}
E_{\nicefrac{1}{2}}^{\rm vac}(d) &= -2\sum_{n~{\rm odd}}\int {d^2k_\perp \over (2\pi)^2} \sqrt{(k^1)^2+(k^2)^2+\left({n\pi \over 2d}\right)^2} \nonumber \\&= -{1\over \pi} \sum_{n~{\rm odd}} \int_{n\pi/2d}^\infty dx\, x^2.
\end{align}
The integral above is divergent and requires regularisation. We adopt a conventional regularisation, which yields the finite vacuum energy
\begin{align}
E_{\nicefrac{1}{2}}^{\rm vac}(d) &= -{1\over \pi} \lim_{\alpha\to 0} {\partial^2 \over \partial \alpha^2} \sum_{n~{\rm odd}} \int_{n\pi/2d}^\infty dx\, e^{-\alpha x} \nonumber \\ &= -{1\over \pi} \lim_{\alpha\to 0} {\partial^2 \over \partial \alpha^2} \left({\pi d\over \alpha^2} -{\pi \over 24d}+{7\pi^3 \alpha^2\over 5760 d^3}+{\mathcal O}(\alpha^3)\right)\nonumber \\ &\approx -{7\pi^2 \over 2880 d^3}.
\end{align}
In the last line above a term linear in $d$, which gives a constant contribution to $\partial E_{\nicefrac{1}{2}}^{\rm vac}(d)/\partial d$ has been ignored. The final result for the Casimir force associated with the massless spin-$\nicefrac{1}{2}$ field between two perfectly reflecting parallel plates is therefore
\begin{align}\label{fincas1}
F_{\nicefrac{1}{2}}(d) = -{\partial E_{\nicefrac{1}{2}}^{\rm vac}(d) \over \partial d} = -{7\pi^2 \over 960 d^4},
\end{align}
which is well-known \cite{chodos_new_1974}.

\subsubsection{Spin-$1$}\label{cas1sec}

For the massless spin-$1$ field in a box with periodic BCs, the vacuum energy according to Eq. (\ref{Hone}) is
\begin{align}\label{evac1}
E_1^{\rm vac} = {1\over 2}\sum_{\bf k} \sum_{\lambda=\pm 1} \omega.
\end{align}
It must be understood in the above expression that when any of the components $k^i$ vanish, there is only one independent polarisation $\lambda$. Substituting into Eq. (\ref{evac1}) the allowed values of ${\bf k}$ given in Eq. (\ref{kone}), and as in the spin-$\nicefrac{1}{2}$ case taking the continuum limit with respect to the $k^1$ and $k^2$-directions, one obtains
\cite{milonni_quantum_1994}
\begin{align}
E_1^{\rm vac}(d) =& \, {\sum_{n=0}^\infty}' \int {d^2k_\perp\over (2\pi)^2} \sqrt{(k^1)^2+(k^2)^2+\left({n\pi \over d}\right)^2} \nonumber \\ =& {1\over 2\pi} {\sum_{n \in {\mathbb N}}}' \int_{n\pi/d}^\infty dx\, x^2
\end{align}
where following \cite{milonni_quantum_1994} we use a prime on the summation to indicate that for the value $n=0$ a factor of $\nicefrac{1}{2}$ must be inserted. The vacuum energy is as expected divergent, and we adopt the same regularisation as in the spin-$\nicefrac{1}{2}$ case. This yields the finite vacuum energy
\begin{align}
E_1^{\rm vac}(d) = {1\over 2\pi} \lim_{\alpha\to 0} {\partial^2 \over \partial \alpha^2} {\sum_{n \in {\mathbb N}}}'\int_{n\pi/d}^\infty dx\, e^{-\alpha x} \approx -{\pi^2 \over 720 d^3}
\end{align}
where as in the spin-$\nicefrac{1}{2}$ case a term linear in $d$ has been ignored. Thus, the final result for the Casimir force associated with the massless spin-$1$ field between two perfectly reflecting parallel plates is
\begin{align}\label{fincas2}
F_1(d) = -{\partial E_1^{\rm vac}(d) \over \partial d} = -{\pi^2 \over 240 d^4},
\end{align}
which is well-known \cite{casimir_attraction_1948}.

The calculations above pertaining to the spin-$\nicefrac{1}{2}$ and spin-$1$ fields reveal that the only ingredients necessary in order to obtain a value for the Casimir force, are an expression for the vacuum energy of the field under consideration, and a set of energy-momentum values allowed by the BCs. 

\subsection{The Casimir effect for spin-$\nicefrac{3}{2}$ and spin-$2$ fields}

To calculate the Casmir force associated with a higher spin field, one must first obtain an expression for the associated vacuum energy. This is much more problematic for higher spin fields than it is for the  fields with spin $s\leq 1$. Higher spin fields possess additional gauge freedom, and the local energy-momentum tensors associated with such fields are generally gauge-dependent. Fortunately the total energy may still be gauge-invariant, and this is the working assumption we will make here.

\subsubsection{spin-$\nicefrac{3}{2}$}

The most common description of massless spin-$\nicefrac{3}{2}$ particles is through the Rarita-Schwinger field \cite{rarita_theory_1941}, which consists of four four-component Dirac bispinors $\psi_\mu$ where the index $\mu$ labels the four distinct bispinors. The Rarita-Schwinger field belongs to the reducible representation ${\mathcal R}:=(\nicefrac{1}{2},\nicefrac{1}{2})\otimes \left[(\nicefrac{1}{2},0) \oplus (0,\nicefrac{1}{2})\right]$, which is build from the two outer spin-$\nicefrac{1}{2}$ irreducible representations and the single inner spin-$1$ irreducible representation of $Sp(2,{\mathbb C})$ \cite{weinberg_quantum_1995} (c.f. Fig. \ref{figreps}). The representation ${\mathcal R}$ has (complex) dimension sixteen, whereas the two outer irreducible representations $(\nicefrac{3}{2},0)$ and $(0,\nicefrac{3}{2})$, that directly correspond to the right and left-helicity states of the physical spin-$\nicefrac{3}{2}$ field, have dimension four (c.f. Fig. \ref{figreps}). This gives a total of eight degrees of freedom, but not all of these are dynamically independent. Gauge-fixing reduces the total number of physical degrees of freedom to four, which is precisely the number required to describe the right and left-helicity states of a massless spin-$\nicefrac{3}{2}$ particle and its anti-particle. The Rarita-Schwinger field contains some twelve redundant degrees of freedom, and accordingly the associated equations of motion are invariant under the gauge transformation
\begin{align}\label{g}
\psi'_\mu = \psi_\mu +\partial_\mu \xi.
\end{align}
Although largely redundant the Rarita-Schwinger field is advantageous, because it can be manipulated in much the same way as the familiar Dirac field for spin-$\nicefrac{1}{2}$ particles. It is for this reason that we use it to obtain an expression for the energy of the massless spin-$\nicefrac{3}{2}$ field.

The Rarita-Schwinger Lagrangian can be written
\begin{align}\label{lag}
{\mathscr L}_{\nicefrac{3}{2}} = -{1\over 2}\epsilon^{\mu\nu\rho\sigma}{\bar \psi}_\mu \gamma^5\gamma_\rho \hspace*{-0.8mm} {\overset{\,\,\text{\tiny $\leftrightarrow$ }}\partial}_{\hspace*{-1mm}\sigma} \psi_\nu
\end{align}
where $f\hspace*{-0.8mm}{\overset{\,\,\text{\tiny $\leftrightarrow$ }}\partial}_{\hspace*{-1mm}\mu}g := f\partial_\mu g - (\partial_\mu f)g$, the $\epsilon^{\mu\nu\rho\sigma}$ are totally anti-symmetric with $\epsilon^{0123}=1$, the set $\{{\gamma^\mu},~\gamma^5\}$ consists of the usual Dirac matrices with $\gamma^5:=i\gamma^0\gamma^1\gamma^2\gamma^3$, and ${\bar \psi}_\mu:= \psi^\dagger_\mu\gamma^0$ denotes the relativistic adjoint of $\psi_\mu$. Under the gauge transformation in Eq. (\ref{g}) the Lagrangian varies by a total divergence.

The twelve redundant degrees of freedom are eliminated using constraints. In particular, the generalised Coulomb gauge is obtained by imposing the constraints \cite{das_gauge_1976}
\begin{align}\label{con}
\gamma^\mu\psi_\mu= &\,0_4, & \partial^\mu\psi_\mu =&\,0_4, & \psi_0 = &\,0_4
\end{align}
where $0_4$ denotes the four-column with all zero entries. Using the identity
\begin{align}
\epsilon^{\mu\nu\rho\sigma}\gamma^5\gamma_\rho \equiv i\left(g^{\mu\nu}\gamma^\sigma -g^{\mu\sigma}\gamma^\nu - g^{\nu\sigma}\gamma^\mu + \gamma^\mu\gamma^\sigma\gamma^\nu\right)
\end{align}
it is a straightforward exercise to verify that for the Lagrangian in Eq. (\ref{lag}) along with the constraints in (\ref{con}), the Euler-Lagrange equations yield the Dirac equation $i\gamma^\mu\partial_\mu \psi_\nu =0_4$ and its adjoint $i(\partial_\mu {\bar \psi}_\nu) \gamma^\mu =0_4^T$. This in turn ensures that $\psi_\mu$ (and its adjoint) satisfy the correct relativistic wave equation
\begin{align}\label{wav}
\square \psi_\mu =0_4
\end{align}
where $\square :=\partial_\mu \partial^\mu$.

The constraints in (\ref{con}) reduce the number of (complex) physical degrees of freedom to the four required in order to describe the right and left-helicity states of a massless spin-$\nicefrac{3}{2}$ particle and its anti-particle. As an ansatz for $\psi_\mu$ in Eq. (\ref{wav}), we therefore make the following Fourier expansion \cite{das_gauge_1976}
\begin{align}\label{mode}
\psi_\mu (x) \hspace*{-0.5mm}= \hspace*{-0.8mm} \int d^3k \hspace*{-0.8mm} \sum_{\lambda = \pm \nicefrac{3}{2}}\hspace*{-0.5mm}\left[b_\lambda({\bf k})u_{\mu}({\bf k},\lambda; x) \hspace*{-0.4mm}+ d^\dagger_\lambda({\bf k})v_{\mu}({\bf k},\lambda; x)\right]
\end{align}
where the $b_\lambda({\bf k}),~b^\dagger_\lambda({\bf k})$ and $d_\lambda({\bf k}),~d^\dagger_\lambda({\bf k})$ are annihilation and creation operators for particles and anti-particles respectively, which as in the spin-$\nicefrac{1}{2}$ case satisfy the anti-commutation relations in Eq. (\ref{acom}). It is of course possible to invert Eq. (\ref{mode}) and its hermitian-conjugate, and this means it is possible to define the $b_\lambda({\bf k}),~b^\dagger_\lambda({\bf k})$ and $d_\lambda({\bf k}),~d^\dagger_\lambda({\bf k})$ in terms of $\psi_\mu(x)$. The single-particle positive and negative-energy wavefunctions $u_\mu({\bf k},\lambda; x)$ and $v_\mu({\bf k},\lambda; x)$ are defined by
\begin{align}
u_\mu({\bf k},\lambda; x):=&{1\over \sqrt{2\omega (2\pi)^3}} u_{\mu}({\bf k},\lambda)e^{-ik_\nu x^\nu}, \nonumber \\ v_\mu({\bf k},\lambda; x):=&{1\over \sqrt{2\omega (2\pi)^3}} v_{\mu}({\bf k},\lambda)e^{ik_\nu x^\nu}
\end{align}
where $u_\mu({\bf k},\lambda)$ and $v_\mu({\bf k},\lambda)$ satisfy the momentum-space counterparts of the constraints in (\ref{con}). The positive and negative-energy wavefunctions are not linearly-independent and are related by $v_\mu({\bf k},\lambda; x)=i\gamma^2\gamma^0 u^\dagger_\mu({\bf k},\lambda; x)$. The orthonormality requirements for $u_\mu({\bf k},\lambda)$ and $v_\mu({\bf k},\lambda)$ can be deduced from the required properties of $u_\mu({\bf k},\lambda; x)$ and $v_\mu({\bf k},\lambda; x)$ under a Lorentz boost, and read \cite{das_gauge_1976}
\begin{align}\label{norm}
&{\bar u}^\mu({\bf k},\lambda)\gamma^\nu u_{\mu}({\bf k},\lambda') ={\bar v}^\mu({\bf k},\lambda)\gamma^\nu v_{\mu}({\bf k},\lambda') = -2k^\nu\delta_{\lambda\lambda'}, \nonumber \\  &{\bar u}^\mu({\bf k},\lambda)\gamma^0 v_{\mu}(-{\bf k},\lambda') =0.
\end{align}

Using the Lagrangian in Eq. (\ref{lag}) and the constraints in (\ref{con}) we obtain the energy-density
\begin{align}
{\mathscr H}_{\nicefrac{3}{2}} ={\partial {\mathscr L}_{\nicefrac{3}{2}}\over \partial (\partial_t \psi^\nu)}\partial_t \psi^\nu - {\mathscr L}_{\nicefrac{3}{2}}=-{i\over 2}{\bar \psi}_\nu \gamma^i \hspace*{-0.8mm}{\overset{\,\,\text{\tiny $\leftrightarrow$ }}\partial}_{\hspace*{-1mm}i} \psi^\nu.
\end{align}
Using this expression together with Eqs. (\ref{acom}), (\ref{mode}) and (\ref{norm}), the total energy can be written as in the spin-$\nicefrac{1}{2}$ case as
\begin{align}\label{H3/2}
H_{\nicefrac{3}{2}}&=\int d^3x \, {\mathscr H}_{\nicefrac{3}{2}} \nonumber \\
&=\int d^3k \sum_{\lambda=\pm \nicefrac{3}{2}} \omega\left(b^\dagger_\lambda({\bf k})b_\lambda({\bf k}) + d^\dagger_\lambda({\bf k})d_\lambda({\bf k}) -1\right).
\end{align}

Since the spin-$\nicefrac{3}{2}$ field is fermionic, according to Eq. (\ref{ks}) the allowed values of ${\bf k}$ for the calculation of the Casimir force, are the same as in the spin-$\nicefrac{1}{2}$ case. Moreover, because the vacuum energies in Eqs (\ref{Hhalf}) and (\ref{H3/2}) are identical, the calculation of the Casimir force from Eq. (\ref{H3/2}) is also the same as in the spin-$\nicefrac{1}{2}$ case, with the final result given by Eq. (\ref{fincas1}).

\subsubsection{Spin-$2$}\label{hspins2}

The spin-$2$ field is most commonly described using a symmetric traceless tensor field $h^{\mu\nu}$, which corresponds to the inner $(1,1)$ irreducible representation of the symplectic group $Sp(2,{\mathbb C})$ [c.f. appendix \ref{A3} and Fig. \ref{figreps}]. This field can also be viewed as the first-order gravitational correction to the components of the Minkowski inner-product. If we expand the general ``metric" tensor of curved spacetime as $g^{\mu\nu}(u)=g^{\mu\nu}+uh^{\mu\nu}+...\,$, then Einstein's vacuum equations yield an equation for $h^{\mu\nu}$ which coincides with the correct relativistic wave equation for a massless spin-$2$ particle (the so-called graviton). As expected, gauge-fixing constraints are required to reduce the number of degrees of freedom present in $h^{\mu\nu}$ to those describing the physical right and left-helicities of the graviton.

These helicities are more directly described by symmetric spin-tensors $\psi^{abcd}$ and $\psi_{{\bar a}{\bar b}{\bar c}{\bar d}}$, which belong to the outer irreducible representations $(2,0)$ and $(0,2)$ respectively, and which satisfy the usual equations of motion (\ref{geneqnmo}). As in the spin-$1$ and spin-$\nicefrac{3}{2}$ cases, it is these fields which belong to outer representations of $Sp(2,{\mathbb C})$, that are in many ways more physically relevant than the gauge-dependent inner field potentials such as $h_{\mu\nu}$ (c.f. Fig. \ref{figreps}).

A significant property of the field $\psi^{abcd}$ is that it can be used to define the so-called Bel-Robinson tensor $T^{\mu\nu\rho\sigma}$, which is one of many possible gravitational versions of a local energy-momentum tensor \cite{penrose_spinors_1987}. 
As expected of an energy-momentum tensor, $T^{\mu\nu\rho\sigma}$ is totally symmetric, traceless and possesses certain positivity properties. It is also the natural analog of the energy-momentum tensor $T^{\mu\nu}$ found in electrodynamics \cite{penrose_spinors_1987}. This last property is most easily seen using the two-spinor calculus formalism, whereby
\begin{align}
T^{\mu\nu\rho\sigma} = {\sigma^\mu}_{{\bar a}a}{\sigma^\nu}_{{\bar b}b}{\sigma^\rho}_{{\bar c}c}{\sigma^\sigma}_{{\bar d}d}\psi^{abcd}\psi^{{\bar a}{\bar b}{\bar c}{\bar d}}.
\end{align}
Thus, the local current $T^{\mu 000}$ is the natural analog of the currents encountered in the spin-$\nicefrac{1}{2}$ and spin-$1$ cases, and as such the generalised BCs in Eq. (\ref{gen}) imply that for the spin-$2$ field, it is the normal component of $T^{\mu000}$ that vanishes at each plate.

In order to obtain an expression for the energy associated with the spin-$2$ field, we build on the early approach of \cite{arnowitt_quantum_1959} for the quantisation of linearised gravity. The starting point in \cite{arnowitt_quantum_1959} is Schwinger's action principle \cite{schwinger_theory_1951}, which is stated in terms of a Lagrangian of an extremely general form. This form includes the so-called Platini general-relativistic Lagrangian as a special case. The use of the Platini-Lagrangian formulation of general relativity turns out to be a crucial feature of the approach adopted in \cite{arnowitt_quantum_1959}, because it circumvents a number of problems pertaining to the highly nonlinear nature of general relativity.

From here on we will employ the shorthand notation $f_{ij,k} := \partial_k\partial f_{ij}$ and $f_{ij,kl} := \partial_l\partial_k f_{ij}$. This allows us to write the linearised (first-order) approximation to the Platini Lagrangian as
\begin{align}\label{lag2}
{\mathscr L}_2=&\,h^{\nu\rho}\left[{\Gamma^\mu}_{\nu\rho,\mu}-{1\over 2}\left({\Gamma^\mu}_{\nu\mu,\rho}+{\Gamma^\mu}_{\rho\mu,\nu}\right)\right] \nonumber \\ &-g^{\nu\rho}\left[{\Gamma^\mu}_{\sigma\nu}\cdot{\Gamma^\sigma}_{\rho\mu}-{\Gamma^\mu}_{\sigma\mu}\cdot{\Gamma^\sigma}_{\nu\rho}\right]
\end{align}
where $2f\cdot g:=fg+gf$, the $h_{\mu\nu}=h_{\nu\mu}$ are the first-order corrections to the components $g_{\mu\nu}$ of the Minkowski inner-product, and the ${\Gamma^\mu}_{\nu\rho} = {\Gamma^\mu}_{\rho\nu}$ are the components of the Levi-Civita connection over spacetime. To first-order in $h_{\mu\nu}$ the connection coefficients can be written
\begin{align}
{\Gamma^\mu}_{\nu\rho} \approx {1\over 2}g^{\mu\sigma}\left(h_{\sigma\nu,\rho}+h_{\sigma\rho,\nu}-h_{\nu\rho,\sigma}\right).
\end{align}
The (constrained) fields $h_{ij}$ and ${\Gamma^0}_{ij}$ serve as canonical fields in the Hamiltonian formulation of linearised gravity given in \cite{arnowitt_quantum_1959}, which we review presently.

The variational principle used in conjunction with the Lagrangian in Eq. (\ref{lag2}) yields the linearised equations of motion \cite{arnowitt_quantum_1959}
\begin{align}\label{eqnmo2}
&\partial_t h_{ij} =-2{\Gamma^0}_{ij} + h_{0i,j}+h_{0j,i}, \nonumber \\
&\partial_t h_{\mu 0} = -\delta_{\mu i}\left(h_{00,i}-{\Gamma^i}_{00}\right)+2\delta^{\mu0}{\Gamma^0}_{00}, \nonumber \\
&\partial_t {\Gamma^0}_{ij} = -{1\over 2}\left(h_{ij,kk}+h_{kk,ij}-h_{ki,jk}-h_{kj,ik}-k_{00,ij}\right)
\end{align}
where it is understood that all repeated indices $i,j,k$ are to be summed over $1,2,3$. The entire theory is invariant under the coordinate (gauge) transformations \cite{arnowitt_quantum_1959}
\begin{align}
&h'_{\mu\nu}=h_{\mu\nu} +\xi_{\nu,\mu} +\xi_{\mu,\nu} - g_{\mu\nu}{\xi^\rho}_{,\rho}, \nonumber \\
&{\Gamma'^\mu}_{\nu\rho} ={\Gamma^\mu}_{\nu\rho}+{\xi^\mu}_{,\nu\rho}
\end{align}
and as such the variational principle also yields the following equations of constraint \cite{arnowitt_quantum_1959}
\begin{align}
&{\Gamma^i}_{jk} = {1\over 2}\left(h_{ij,k}+h_{ik,j}+h_{jk,i}\right),\nonumber \\
&{\Gamma^\mu}_{i0} = \delta^{\mu k}\left(-h_{0k,i}+{\Gamma^0}_{ki}\right)+{1\over 2}\delta^{\mu 0}h_{00,i}, \nonumber \\
&{\Gamma^0}_{ik,k}={\Gamma^0}_{kk,i},~~~~~~h_{ij,ij}=h_{ii,jj}.
\end{align}
The constraints arise due to the presence of redundant gauge degrees of freedom, which must be eliminated via gauge fixing. A common choice of gauge in general relativity is the transverse traceless (TT) gauge, which is analogous to the Coulomb gauge in electrodynamics. In the TT-gauge the field $h_{\mu\nu}$ satisfies
\begin{align}\label{tt}
{h^{\mu\nu}}_{,\mu} =&\,0, & {h^\mu}_\mu =&\,0, & h_{\mu 0}=&\,0,
\end{align}
and the equations of motion (\ref{eqnmo2}) take on the following simple harmonic form, which is synonymous with the free dynamics generated by quadratic Hamiltonians;
\begin{align}\label{eqnmo2b}
\partial_t h_{ij} &= -2{\Gamma^0}_{ij} & \partial_t {\Gamma^0}_{ij} &= -{1\over 2} h_{ij,kk}.
\end{align}
Together the constraint $h_{\mu0} =0$ and the equations of motion (\ref{eqnmo2b}) ensure the correct relativistic wave equation
\begin{align}\label{wave}
\square h_{\mu\nu} = 0.
\end{align}

In order to proceed it will be helpful to count the number of physical degrees of freedom with which we are dealing. The TT-gauge constraints in (\ref{tt}) reduces the number of physical degrees of freedom in $h_{\mu\nu}$ from ten to two. To see this note that there are nine separate equations in (\ref{tt}), but that only eight of these are independent, because the first constraint with $\nu=0$ is implied by the last. Thus, in total, we are left with four real degrees of freedom, two of which are contained in $h_{ij}$, and two of which are contained in $\partial_t h_{ij} =-2{\Gamma^0}_{ij}$. This is precisely the number of physical degrees of freedom required to describe the two independent helicities of the graviton. As an ansatz for $h_{\mu\nu}$ in Eq. (\ref{wave}), we therefore make the following Fourier expansion
\begin{align}\label{h}
h_{\mu\nu} (x) =& \nonumber \\ \int d^3k &\sum_{\lambda =\pm 2}\left[u_{\mu\nu}({\bf k},\lambda; x) a_\lambda({\bf k}) + {\bar u}_{\mu\nu}({\bf k},\lambda; x) a^\dagger_\lambda({\bf k})\right]
%
\end{align}
where the single-particle wavefunctions $u^{\mu\nu}({\bf k},\lambda; x)$ are defined by
\begin{align}\label{singpart2}
u_{\mu\nu}({\bf k},\lambda; x) := {1\over \sqrt{2\omega(2\pi)^3} }e_{\mu\nu}({\bf k},\lambda)e^{-ik_\rho x^\rho}
\end{align}
in which $e_{\mu\nu}({\bf k},\lambda)$ denotes a transverse traceless polarisation tensor. In Eq. (\ref{h}) we have used $a^\dagger_\lambda({\bf k})$ to denote the complex-conjugate of the complex number $a_\lambda({\bf k})$ in anticipation of the quantum mechanical expression for $h_{\mu\nu}$.

The polarisation tensors $e_{\mu\nu}({\bf k},\lambda)$ in Eq. (\ref{singpart2}) possess the orthonormality and symmetry properties
\begin{align}\label{enorm}
&e_{\mu\nu}({\bf k},\lambda){\bar e}^{\mu\nu}({\bf k},\lambda') =2\delta_{\lambda\lambda'}, & g^{\mu\nu} &e_{\mu\nu}({\bf k},\lambda)=0, \nonumber \\
&e_{ij}(-{\bf k},\lambda)	= e_{ij}({\bf k},\lambda), & e_{i0}(-{\bf k},\lambda)	&= -e_{i0}({\bf k},\lambda),
\end{align}
which can be deduced from the required properties of $u_{\mu\nu}$ under a Lorentz boost. If we now define the following inner-product over the Hilbert space of solutions to Eq. (\ref{wave})
\begin{align}
\langle h , h'\rangle := {i\over 2}\int d^3 x \, {\bar h}^{\mu\nu}(x)\hspace*{-1mm}{\overset{\,\,\text{\tiny $\leftrightarrow$ }}\partial}_{\hspace*{-1mm}t}h'_{\mu\nu}(x)\bigg|_{t=0},
\end{align}
we see that the normalisation of the polarisation tensors $e^{\mu\nu}({\bf k},\lambda)$ has been chosen such that the wavefunctions $u^{\mu\nu}({\bf k},\lambda; x)$ satisfy the orthonormality conditions
\begin{align}
\langle u({\bf k},\lambda),u({\bf k}',\lambda') \rangle &= \delta_{\lambda\lambda'}\delta({\bf k}-{\bf k}'), \nonumber \\  \langle {\bar u}({\bf k},\lambda),{\bar u}({\bf k}',\lambda') \rangle &= -\delta_{\lambda\lambda'}\delta({\bf k}-{\bf k}').
\end{align}

The mode expansion in Eq. (\ref{h}) is consistent with the TT-gauge constraints in (\ref{tt}) provided we impose the following further restrictions on the polarisation tensors
\begin{align}\label{enormb}
k^\mu e_{\mu\nu}({\bf k},\lambda)=&\,0, & {e^\mu}_{\mu}({\bf k},\lambda) =&\, 0, & e_{\mu0}({\bf k},\lambda) =&\, 0.
\end{align}
The first two constraints above are consistent with the required behaviour of $h_{\mu\nu}$ under a Lorentz boost. The third however, clearly shows that $h_{\mu\nu}$ cannot transform in a Lorentz-covariant way. This situation is analogous to the one encountered in electrodynamics. Like the components $A_\mu$ of the electromagnetic four-potential, the components $h_{\mu\nu}$ are to be viewed as the components of a geometric field that is physically invariant under a larger transformation group including both Lorentz transformations and gauge transformations. The constraint $h_{\mu0} =0$ is consistent with this broader notion of physical invariance.

Using now Eq. (\ref{eqnmo2b}), we obtain from $h_{\mu\nu}$ in Eq. (\ref{h}) the following expression for ${\Gamma^0}_{ij}$, which is the (negative of the) canonical momentum conjugate to $h_{ij}$
\begin{align}\label{gam}
{\Gamma^0}_{ij} (x) =& {i\over 2} \int d^3k\, \sqrt{\omega \over 2 (2\pi)^3} \sum_{\lambda = \pm 2}\nonumber \\ & \hspace*{-6mm}\times \left(a_\lambda({\bf k})e_{ij}({\bf k},\lambda)e^{-ik_\rho x^\rho} \hspace*{-0.8mm}- a^\dagger_\lambda({\bf k}){\bar e}_{ij}({\bf k},\lambda)e^{ik_\rho x^\rho}\right).
\end{align}
It is of course possible to invert the expressions for $h_{ij}$ [in Eq. (\ref{h})] and ${\Gamma^0}_{ij}$ given above. This means the $a_\lambda({\bf k})$ and $a^\dagger_\lambda({\bf k})$ can be defined in terms of the Fourier transforms of the canonical fields.

The energy-density can be calculated from the Lagrangian in Eq. (\ref{lag2}) and after making use of the constraints (\ref{tt}) can be written (in the TT-gauge) \cite{arnowitt_quantum_1959}
\begin{align}
{\mathscr H}_2 ={\partial {\mathscr L}_2\over \partial (\partial_t h^{\mu\nu})}\partial_t h^{\mu\nu} - {\mathscr L}_2=({\Gamma^0}_{ij})^2 + {1\over 4}(h_{ij,k})^2
\end{align}
where the indices $i, j, k$ are each summed over $1,2,3$. This yields the total energy \cite{arnowitt_quantum_1959}
\begin{align}\label{H2}
H_2=\int d^3 x \, {\mathscr H}_2 = \int d^3 x\, \left[ ({\Gamma^0}_{ij})^2 + {1\over 4}(h_{ij,k})^2 \right].
\end{align}

When the fields $h_{ij}$ and ${\Gamma^0}_{ij}$ are interpreted as operators on some suitably defined Hilbert space, the above Hamiltonian together with the Heisenberg equation $i\partial_t {\mathcal O} =[{\mathcal O},H_2]$ yields the correct equations of motion (\ref{eqnmo2b}), provided the canonical fields satisfy the following equal-time commutation relation, consistent with the TT-gauge constraints \cite{arnowitt_quantum_1959}
\begin{align}\label{com2}
[h_{ij}({\bf x}),&{\Gamma^0}_{kl}({\bf x}')]=-i\delta_{ijkl}^{\rm TT}({\bf x}-{\bf x}')\nonumber \\ &:={i\over 2}\left[\left({2\over 3}\delta_{ij}\delta_{kl} - \delta_{ik}\delta_{jl} - \delta_{il}\delta_{jk}\right)\delta({\bf x}-{\bf x}')\right]^{\rm TT}.
\end{align}
The right-hand-side of the above equality attains meaning when the expression within the brackets is contracted with a suitably chosen three-dimensional symmetric test tensor $f_{kl}({\bf x}')$ and is then integrated over a fixed time-slice. This procedure will yield some symmetric tensor $g_{ij}({\bf x})$, which in general, will not be transverse or traceless. The notation $[\cdot]^{\rm TT}$ signifies that the transverse traceless component of $g_{ij}({\bf x})$ must then be taken, and it is this projection onto the TT subspace, which ensures Eq. (\ref{com2}) is consistent with the TT-gauge constraints in (\ref{tt}). In short, the distribution $\delta_{ijkl}^{\rm TT}({\bf x}-{\bf x}')$ is defined by the integration condition
\begin{align}\label{intcon}
\int d^3x' \, \delta_{ijkl}^{\rm TT}({\bf x}-{\bf x}')f_{ij}({\bf x}') = f^{\rm TT}_{kl}({\bf x})
\end{align}
where $f^{\rm TT}_{kl}({\bf x})$ is the transverse traceless component of the symmetric tensor $f_{kl}({\bf x})$. This situation is of course quite familiar from quantum electrodynamics in the Coulomb gauge whereby the right-hand-side of the canonical commutation relation that is analogous to Eq. (\ref{com2}), involves the transverse delta function $\delta_{ij}^{\rm T}({\bf x}-{\bf x}')$ \cite{cohen-tannoudji_photons_1989}. In the spin-$2$ case however, one is dealing with three-dimensional symmetric tensors rather than three-dimensional vectors, and the appropriate distribution is therefore $\delta_{ijkl}^{\rm TT}({\bf x}-{\bf x}')$. In terms of an integral in momentum-space $\delta_{ijkl}^{\rm TT}({\bf x}-{\bf x}')$ admits the representation
\begin{align}\label{com3}
\delta_{ijkl}^{\rm TT}({\bf x}-{\bf x}') \hspace*{-0.5mm}=\hspace*{-0.5mm}\int \hspace*{-1mm}{d^3k\over 2(2\pi)^3} \hspace*{-1mm}\sum_{\lambda= \pm 2} \hspace*{-1mm}e_{ij}({\bf k},\lambda){\bar e}_{kl}({\bf k},\lambda)e^{i{\bf k}\cdot ({\bf x}-{\bf x}')},
\end{align}
which is also reminiscent of the well-known Fourier transform representation of the transverse delta function $\delta_{ij}^{\rm T}({\bf x}-{\bf x}')$ \cite{craig_molecular_2012}.

To verify that the expression on the right-hand-side of Eq. (\ref{com3}) possesses the property specified in Eq. (\ref{intcon}) we first note that any (suitably well-behaved) real transverse traceless symmetric tensor can be decomposed as
\begin{align}\label{ttf}
f_{ij}^{\rm TT}({\bf x})  &=\int {d^3 k \over \sqrt{(2\pi)^3}} {\tilde f}_{ij}^{\rm TT}({\bf k}) e^{i{\bf k}\cdot {\bf x}} 
\nonumber \\&=\int {d^3 k \over \sqrt{(2\pi)^3}} \sum_{\lambda =\pm 2} e_{ij}({\bf k},\lambda) {\tilde f}_\lambda({\bf k})e^{i{\bf k}\cdot {\bf x}}
\end{align}
where ${\tilde f}_{ij}^{\rm TT}$ denotes the (three-dimensional) Fourier transform of $f_{ij}^{\rm TT}$, and ${\tilde f}_\lambda({\bf k})$ denotes a complex Fourier coefficient labeled by helicity $\lambda$ and momentum ${\bf k}$. We also make use of the following more complicated scalar-vector-tensor decomposition of a (suitably well-behaved) real symmetric three-dimensional tensor
\begin{align}\label{arbdec}
f_{ij}({\bf x}) =\int {d^3 k \over \sqrt{(2\pi)^3}} {\tilde f}_{ij}({\bf k}) e^{i{\bf k}\cdot {\bf x}} 
\end{align}
with \cite{vilenkin_cosmic_2000}
\begin{align}\label{arbdec2}
{\tilde f}_{ij} = {1\over 3}{\tilde f}\delta_{ij} + \left({\hat k}_i{\hat k}_j -{1\over 3}\delta_{ij}\right){\tilde f}^s + {\hat k}_i{\tilde f}^v_j +  {\hat k}_j{\tilde f}^v_i +{\tilde f}_{ij}^{\rm TT}.
\end{align}
Here ${\hat {\bf k}}:={\bf k}/|{\bf k}|$ and ${\tilde f}$ denotes the trace of ${\tilde f}_{ij}$. The scalar quantity ${\tilde f}^s$ and the vector quantity ${\tilde {\bf f}}^v$ are defined in terms of the ${\hat k}_i$, ${\tilde f}_{ij}$ and ${\tilde f}$, but their precise form need not be given in order to prove that Eq. (\ref{com3}) holds. Now, due to the constraints in (\ref{enormb}), we have using Eqs. (\ref{ttf}) and (\ref{arbdec2}) that
\begin{align}\label{73}
\sum_{\lambda=\pm 2} e_{ij}({\bf k},\lambda){\bar e}_{kl}({\bf k},\lambda){\bar {\tilde f}}_{ij}({\bf k}) = {\bar {\tilde f}}_{kl}^{\rm TT}({\bf k}).
\end{align}
Thus, substituting Eq. (\ref{com3}) and the complex-conjugate of Eq. (\ref{arbdec}) into the left-hand-side of Eq. (\ref{intcon}), one readily obtains by making use of Eq. (\ref{73}), the complex-conjugate of the right-hand-side of Eq. (\ref{ttf}). This proves that Eq. (\ref{com3}) holds when used in conjunction with any (suitably well-behaved) real symmetric tensor. 

Having established that Eq. (\ref{com3}) holds we can now check that our canonical formulation of linearised gravity is consistent with the mode expansions in Eqs. (\ref{h}) and (\ref{gam}). Consistency is ensured if when the mode expansions are substituted into the commutation relation $i[h_{ij}({\bf x}),{\Gamma^0}_{kl}({\bf x}')]$, one obtains the right-hand-side of Eq. (\ref{com3}). This will be the case provided the $a_\lambda({\bf k})$ and $a^\dagger_\lambda({\bf k})$ obey the bosonic commutation relation in Eq. (\ref{com}). Thus, imposing the bosonic commutation relation between $a_\lambda({\bf k})$ and $a^\dagger_\lambda({\bf k})$, and substituting Eqs. (\ref{h}) and (\ref{gam}) into Eq. (\ref{H2}), we obtain using Eq. (\ref{enorm}) the expected expression for the energy of the massless spin-$2$ field;
\begin{align}\label{H22}
H_2=\int d^3k \sum_{\lambda =\pm 2} \omega \left(a^\dagger_\lambda({\bf k})a_\lambda({\bf k}) +{1\over 2} \right).
\end{align}
This Hamiltonian together with the Heisenberg equation yields the correct equations of motion (\ref{eqnmo2}).

Since the spin-$2$ field is bosonic, according to Eq. (\ref{ks}) the allowed values of ${\bf k}$ for the calculation of the Casimir force, are the same as in the spin-$1$ case. Moreover, because the vacuum energies in Eqs (\ref{Hone}) and (\ref{H22}) are identical, the calculation of the Casimir force from Eq. (\ref{H22}) is also the same as in the spin-$1$ case, with the final result given by Eq. (\ref{fincas2}).

Having obtained the Casimir forces associated with both the spin-$\nicefrac{3}{2}$ and spin-$2$ fields, we are in a position to consider what might be called the super-gravitational Casimir force. In supergravity the graviton is paired with its supersymmetric partner---the spin-$\nicefrac{3}{2}$ gravitino. The Lagrangian can be taken as a sum of uncoupled Lagrangians associated with the spin-$\nicefrac{3}{2}$ and spin-$2$ fields respectively, plus a term involving additional auxiliary fields. In the case of pure gravitation these auxiliary fields vanish on the space of solutions to the equations of motion \cite{weinberg_quantum_2000}. We can therefore conjecture that the super-gravitational Casimir force is the sum of the fermionic and bosonic Casimir forces given in Eqs. (\ref{fincas1}) and (\ref{fincas2}) respectively. 

\section{Conclusions}

In this paper we have used general physical BCs to calculate the Casimir force between two perfectly reflecting parallel plates for the massless quantum fields up to spin-$2$. For each spin value the generalised BCs imply that at the plates the normal component of a physically appropriate local-current vanishes. For the spin-$\nicefrac{1}{2}$ (massless Dirac-Weyl) field the appropriate current is the particle-current. For the spin-$1$ (Maxwell) field no particle-current exists, so the electromagnetic energy-current occurs in its place. For the spin-$2$ (linearised gravitational) field neither a local particle-current nor a local energy-current exists, so a current defined in terms of the Bel-Robinson tensor occurs instead.

We have shown that the generalised BCs imply that the allowed values of energy-momentum between two perfectly reflecting parallel plates are the same for all fermionic fields and the same for all bosonic fields. We have verified that these allowed values of energy-momentum lead to two distinct Casimir forces, one associated with fermions and one associated with bosons. This has been achieved through the explicit calculation of the Casimir forces associated with the fields ranging from spin-$\nicefrac{1}{2}$ up to spin-$2$. A significant implication of these general results is that periodic BCs cannot be applied to a fermionic field confined between two parallel plates. This renders certain previous investigations unphysical.

The results we have obtained open up numerous avenues for further investigation into Casimir forces, both for the more familiar spin-$\nicefrac{1}{2}$ and spin-$1$ fields, and for higher-spin fields as well. An obvious extension of the present work lies in the diversification of the surface geometries and reflection properties assumed in the calculations. We have also already mentioned at the end of section \ref{hspins2} a further possible extension of the present work into the arena of supersymmetric field theories. Yet another extension lies in the consideration of confined interacting massless fields, such as coupled spin-$\nicefrac{1}{2}$ and spin-$1$ fields. In this case, in order to determine the allowed values of energy-momentum, one would need to find modified single-particle solutions to the coupled equations of motion, which satisfy the physical constraints of giving no local spin-$\nicefrac{1}{2}$ or spin-$1$ current normal to the surfaces.
\\

{\em Acknowledgements}: The authors would like to thank Almut Beige and Giandomenico Palumbo for several useful discussions regarding this work. This work was supported in part by the UK Engineering and Physical Sciences Research Council (EPSRC). 

\begin{appendix}
\section{Two-spinor calculus}\label{A}

\subsection{Elementary symplectic spinor spaces}

The two-spinor calculus provides a means by which to build arbitrary irreducible representations of the (proper orthochronus) Lorentz group ${\mathcal L}_+^\uparrow$. Representations up to a phase of ${\mathcal L}_+^\uparrow$ are in one-to-one correspondence with representations of its universal covering group $SL(2,{\mathbb C})$, which is identical to the complex symplectic group $Sp(2,{\mathbb C})$. Spinors are built using the two-dimensional complex symplectic vector spaces $S$ and ${\bar S}$ where a bar is used to denote the complex-conjugate space. The space $S$ is the pair $(V,\omega)$, where $V$ is a two-dimensional complex vector space and $\omega:V\times V\to {\mathbb C}$ is a complex symplectic (non-degenerate) form. Choosing a basis $\{f_a\}\subset V$ and employing the summation convention for repeated upper and lower indices we have
\begin{align}
\psi = \psi^a f_a \in S,~~~~~~{\bar \psi}=\psi^{\bar a}f_{\bar a}\in {\bar S}
\end{align}
where we use bars rather than the more commonly used dots to distinguish between a spinor index and a conjugate-spinor index. Furthermore we rely entirely on the different indices in order to distinguish between the components of $\psi$ and ${\bar \psi}$ as well as between the basis vectors $\{f_a\}$ and their conjugates. With these index conventions matrix operations become particularly simple. If a matrix $v$ has elements $v^{ab}$ then we have the following representations
\begin{align}
v\leftrightarrow v^{ab},~~~~{\bar v} \leftrightarrow v^{{\bar a}{\bar b}},~~~~{v}^T \leftrightarrow v^{ba},~~~~~{v}^\dagger \leftrightarrow v^{{\bar b}{\bar a}}
\end{align}
where $^T$ and $^\dagger$ denote matrix transposition and hermitian conjugation respectively. A hermitian matrix clearly has components $v^{a{\bar a}}$ (or $v^{{\bar a}a}$).

In order to construct arbitrary spinors one also uses the dual spaces $S^*, {\bar S}^*$. The dual $V^*$ of the (complex) linear space $V$ is defined as the space of linear maps from $V$ to ${\mathbb C}$. Given a basis $\{e_\mu\} \subset V$, the corresponding dual basis $\{e_\mu\} \subset V^*$ is defined by the orthonormality condition $e^\mu(e_\nu) :=\delta^\mu_\nu$. Every finite-dimensional linear space is isomorphic to its dual, because the two spaces necessarily have the same dimension. However, the identification of $V$ and $V^*$ is not canonical, and one has considerable freedom in pairing vectors with dual vectors using some form of bilinear mapping. In Minkowski spacetime $E^{1,3}$ for example, the map used is a symmetric bilinear inner-product $g:E^{1,3}\times E^{1,3}\to {\mathbb R}$, which with respect to some basis $\{e_\mu\}$ can be written $g=g_{\mu\nu}e^\mu \otimes e^\nu$. For a given vector $v=v^\mu e_\mu \in E^{1,3}$ one defines the corresponding dual vector $v^* := v_\mu e^\mu \in {E^{1,3}}^*$ such that $v_\mu := g_{\mu\nu}v^\nu$. This pairing means that the contraction of a vector $v$ with its dual $v^*$ gives the $g$-norm of $v$, i.e., $v^*(v) \equiv g(v,v)$. In the two-spinor calculus formalism the analogous vector-dual vector pairing is defined with respect to the symplectic structures on $S$ and ${\bar S}$. The dual of an element $\psi \in S$ is an element $\psi^*:S\to {\mathbb C}$ belonging to $S^*$ such that
\begin{align}
\psi^*(\phi) = \omega(\psi,\phi).
\end{align}
Obviously the same relation holds between elements of ${\bar S}$ and ${\bar S}^*$. The symplectic maps $\omega^*:V^*\times V^*\to {\mathbb C}$ and ${\bar \omega}^*:{\bar V}^*\times {\bar V}^*\to {\mathbb C}$ associated with the dual spaces are non-degenerate bivectors. To build spinors one uses distinguished canonical (a.k.a. Darboux, a.k.a. spin) bases $\{f_a\}$,~$\{f_{\bar a}\}$,~$\{f^a\}$ and $\{f^{\bar a}\}$, which belong in $V,~{\bar V},~V^*$ and ${\bar V}^*$ respectively, and which satisfy the relations $f^a(f_b) = f_b(f^a)= \delta^a_b$ and $f^{\bar a}(f_{\bar b}) =f_{\bar b}(f^{\bar a}) = \delta^{\bar a}_{\bar b}$. In the canonical bases the symplectic maps $\omega,~{\bar \omega},~\omega^*$ and ${\bar \omega}^*$ can be written
\begin{align}
&\omega = \omega_{ab}f^a \wedge f^b, ~~~~~~{\bar \omega} = \omega_{{\bar a}{\bar b}}f^{\bar a} \wedge f^{\bar b},\nonumber \\
& \omega^*=\omega^{*ab}f_a\wedge f_b,~~~\, {\bar \omega}^*=\omega^{*{\bar a}{\bar b}}f_{\bar a}\wedge f_{\bar b}.
\end{align}
where $\wedge$ denotes the exterior product. The sets of components $\omega_{ab},~\omega_{{\bar a}{\bar b}},~\omega^{*ab}$ and ${\bar \omega}^{*{\bar a}{\bar b}}$  each support the matrix representation
\begin{align}\label{symmatrep}
\omega_{ab},~\omega_{{\bar a}{\bar b}},~\omega^{*ab},~\omega^{*{\bar a}{\bar b}} \leftrightarrow \left( {\begin{array}{cc}
0 & 1  \\
 -1 & 0  \\
 \end{array} } \right).
\end{align}
As we remarked above $Sp(2,V)$ is the symmetry group associated with $\omega$, that is, given any $T\in Sp(2,V)$ we have that $\omega(T\psi,T\phi)=\omega(\psi,\phi)$ for all $\psi,\phi\in V$. In components this condition along with the analogous conditions pertaining to ${\bar \omega},~\omega^*$ and ${\bar \omega}^*$, can be written
\begin{align}
&{T^a}_b\omega^{bc}{T^d}_c = \omega^{ad},~~~~{T^{\bar a}}_{\bar b}\omega^{{\bar b}{\bar c}}{T^{\bar d}}_{\bar c}=\omega^{{\bar a}{\bar d}},\nonumber \\ &{{\tilde T}_a}^{~ b}\omega_{bc}{\tilde T}_d^{~ c} = \omega_{ad},~~~~~{{\tilde T}_{\bar a}}^{~\bar b}\omega_{{\bar b}{\bar c}}{{\tilde T}_{\bar d}}^{~\bar c}=\omega_{{\bar a}{\bar d}}
\end{align}
where a tilde has been used to denote the matrix contragradient ${\tilde T}:=(T^{T})^{-1}$.

\subsection{Lorentz transformations and spinor index gymnastics}\label{Lts}

We have now all of the ingredients necessary in order to define arbitrary higher order spinors (spin-tensors) and to be able to perform spinor index gymnastics. Denoting the $r$-times cartesian product of a set $S$ with itself by $S^r$, we define a spin-tensor of type-${p~q\choose r~s}$ as a multilinear map $\psi:S^{*p}\times{\bar S}^{*q}\times S^r \times {\bar S}^s \to {\mathbb C}$, which in the canonical bases can be written
\begin{align}
\psi =& {{{\psi^{a_1...a_p}}_{b_1...b_r}}^{{\bar a}_1...{\bar a}_q}}_{{\bar b}_1...{\bar b}_s}f_{a_1}\otimes...\otimes f_{a_p} \nonumber \\ &\otimes f^{b_1}\otimes...\otimes f^{b_r} \otimes f_{{\bar a}_1}\otimes...\otimes f_{{\bar a}_q}\otimes f^{{\bar b}_1}\otimes...\otimes f^{{\bar b}_s}.
\end{align}
When viewed as a spin-tensor field over Minkowski spacetime, the components of $\psi$ are to be viewed as functions of $x \in E^{1,3}$.

A Lorentz transformation $\Lambda \in {\mathcal L}^\uparrow_+$ has two elementary two-dimensional nonequivalent complex representations, one on $S$, which is denoted $T(\Lambda)$, and the complex-conjugate representation on ${\bar S}$, which is denoted ${\bar T}(\Lambda)$. These representations correspond to left-helicity and right-helicity two-spinors respectively. The representation ${\tilde T}(\Lambda)$ on $S^*$ is equivalent to $T(\Lambda)$ on $S$, while the representation ${\tilde {\bar T}}(\Lambda)$ on ${\bar S}^*$ is equivalent to ${\bar T}(\Lambda)$ on ${\bar S}$. The carrier space of the $(i,j)$'th irreducible representation of $Sp(2,{\mathbb C})$ can be taken as the collection of spin-tensors with components $\psi^{(a_1...a_i)({\bar a}_1...{\bar a}_j)}$, where the notation $\psi^{(a_1...a_n)}$ means that $\psi$ is totally-symmetric in the indices $a_1,...,a_n$. In this case the carrier space of the $(i,j)$'th representation has clearly been built out of the spaces $S$ and ${\bar S}$. Alternatively we could, in the obvious way, take any one of the three elementary combinations $S^*$ and ${\bar S}^*$, or $S$ and ${\bar S}^*$, or ${\bar S}$ and $S^*$, as the building blocks for the carrier spaces of the various $(i,j)$ representations. In this paper we choose the pair $(S,{\bar S}^*)$, which means that right-helicity spinors are labelled by unbarred upper indices while left-helicity spinors are labelled by barred lower indices.

Under a Lorentz transformation $\Lambda \in {\mathcal L}^\uparrow_+$ each unbarred upper index of the components of a spin-tensor ${{{\psi^{a...}}_{...}}^{...}}_{...}$ transforms through multiplication by a matrix ${T^b}_a(\Lambda)$. An unbarred lower index ${{{\psi^{...}}_{a...}}^{...}}_{...}$ transforms through multiplication by ${\tilde T}_a^{~b}(\Lambda)$. Obviously the barred indices transform in the same way through matrices ${T^{\bar b}}_{\bar a}(\Lambda)$ and ${\tilde T}_{\bar a}^{~{\bar b}}(\Lambda)$. In the case of a spin-tensor field the argument $x \in E^{1,3}$ transforms to $\Lambda^{-1}x$. More generally under a Poincar\'e transformation $(\Lambda,a)$ with $(\Lambda,a)x = \Lambda x +a$, the argument $x$ transforms as $(\Lambda,a)^{-1}x=\Lambda^{-1}x - \Lambda^{-1}a$.

To carry out spinor index gymnastics the symplectic maps are used, because these are the maps we have used to define the dual spaces. We use $\omega$ to denote the components of these maps in keeping with the common convention in symplectic geometry. The symbol $\epsilon$ is more commonly found in the two-spinor calculus literature. Because the components of the symplectic maps are anti-symmetric one must be careful in raising and lowering indices, for example $\omega^{ab}\psi_a =-\psi_b \neq \psi_b$. We adopt the convention that $\omega_{ab}$ can only be used to lower an index when the repeated index is in the first slot. Similarly $\omega^{ab}$ only raises the index when the repeated index is in the second slot. The same rules apply for barred indices, so altogether
\begin{align}
\omega_{ab}\psi^a=\psi_b,~~\omega^{ab}\psi_b=\psi^a,~~\omega_{{\bar a}{\bar b}}\psi^{\bar a}=\psi_{\bar b},~~\omega^{{\bar a}{\bar b}}\psi_{\bar b} = \psi^{\bar a}.
\end{align}
In what follows we will specify how spinor index gymnastics relates to the usual spacetime index gymnastics.

\subsection{Spinor indices and spacetime indices}\label{A3}

Minkowski spacetime $E^{1,3}$ will be viewed as a pair $(R,g)$ where $R$ is a real four-dimensional vector space and $g:R\times R\to {\mathbb R}$ is a symmetric indefinite inner-product. In an orthonormal basis $\{e_\mu\}$ of $R$ with dual basis $\{e^\mu\}\subset R^*$ (s.t $e^\mu(e_\nu)=\delta^\mu_\nu$) the inner-product $g$ can be written $g=g_{\mu\nu}e^\mu e^\nu$, where the juxtaposition $e^\mu e^\nu$ denotes the symmetric tensor product of $e^\mu$ and $e^{\nu}$. Similarly one defines the inner-product $g^* = g^{\mu\nu}e_\mu e_\nu$ on $R^*$. The components $g_{\mu\nu}$ and $g^{\mu\nu}$ have the matrix representation ${\rm diag}(1,-1,-1,-1)$ and can be used to raise and lower spacetime indices in the usual way; $g^{\mu\nu}v_\nu = v^\mu$ etc.

To relate spinor and spacetime indices we use a representation $[~]:E^{1,3} (E^{1,3*}) \to H(2,{\mathbb C})$ of vectors (dual-vectors) $v\in E^{1,3}~(v^*\in E^{1,3*})$ as two-dimensional hermitian matrices $[v]~([v^*])\in H(2,{\mathbb C})$. The representation $[\cdot ]$ is defined by
\begin{align}
[v]:={\tilde \sigma}^\mu v_\mu ={\tilde \sigma}_\mu v^\mu,~~~~~~[v^*] := \sigma^\mu v_\mu =\sigma_\mu v^\mu
\end{align}
where $\sigma_\mu$ and ${\tilde \sigma}_\mu$ denote the following normalised Pauli spin-matrices
\begin{align}\label{Pauli}
&\sigma^0 ={\tilde \sigma}^0 ={1\over{\sqrt 2}}\hspace*{-0.7mm}\left( {\begin{array}{cc}
1 & 0  \\
 0 & 1  \\
 \end{array} } \right),~~~~~\sigma^1 =-{\tilde \sigma}^1 ={1\over {\sqrt 2}}\hspace*{-0.7mm}\left( {\begin{array}{cc}
0 & 1  \\
 1 & 0  \\
 \end{array} } \right),\nonumber \\
&\sigma^2 =-{\tilde \sigma}^2 ={1\over {\sqrt 2}}\hspace*{-0.7mm}\left( {\begin{array}{cc}
0 & -i  \\
 i & 0  \\
 \end{array} } \right),~\sigma^3 =-{\tilde \sigma}^3 ={1\over {\sqrt 2}}\hspace*{-0.7mm}\left( {\begin{array}{cc}
1 & 0  \\
 0 & -1  \\
 \end{array} } \right).
\end{align}
The components of the Minkowski inner-product have numerous representations in terms of the normalised Pauli spin-matrices;
\begin{align}
{\rm tr}(\sigma_\mu {\tilde \sigma}_\nu) \hspace*{-0.3mm}=\hspace*{-0.3mm} {\rm tr}({\tilde \sigma}_\mu \sigma_\nu) \hspace*{-0.3mm}=\hspace*{-0.3mm} g_{\mu\nu},~\,
{\rm tr}(\sigma^\mu {\tilde \sigma}^\nu) \hspace*{-0.3mm}=\hspace*{-0.3mm} {\rm tr}({\tilde \sigma}^\mu \sigma^\nu) \hspace*{-0.3mm}=\hspace*{-0.3mm} g^{\mu\nu},
\end{align}
which enables one to define the Minkowski inner-product on $H(2,{\mathbb C})$ as
\begin{align}
g(v,w) &= g_{\mu\nu}v^\mu w^\nu ={\rm tr}(\sigma_\mu {\tilde \sigma}_\nu)v^\mu w^\nu \nonumber \\ &= {\rm tr}([v^*][w]) =: [v]\cdot [w].
\end{align}
With the elements of the matrices $[v]$ and $[v^*]$ we can associate components of spin-tensors via
\begin{align}\label{sv}
&v_{{\bar a}a} = {\sigma^\mu}_{{\bar a}a}v_\mu = \sigma_{\mu {\bar a}a}v^\mu,~~~~~~v^{a{\bar a}} = {\tilde \sigma}^{\mu a{\bar a}}v_\mu ={{\tilde \sigma}_\mu}^{~\, a{\bar a}}v^\mu \nonumber \\ &v^\mu = {\sigma^\mu}_{{\bar a}a}v^{a{\bar a}} = {\tilde \sigma}^{\mu a{\bar a}}v_{{\bar a}a},~~~~~v_\mu = \sigma_{\mu{\bar a}a}v^{a{\bar a}} ={{\tilde \sigma}_\mu}^{~\, a{\bar a}}v_{{\bar a}a}.
\end{align}
The Minkowski inner-product can be written in terms of spin-tensor components in numerous ways, for example $
g(v,w) = v^{a{\bar a}}w_{{\bar a}a}$. The hybrid so-called van der Waerden symbols are defined as
\begin{align}\label{vdw}
&{g_\mu}^{a{\bar a}} :=  {\sigma_\mu}^{a{\bar a}},~~~~~~~~~{g^\mu}^{a{\bar a}} :=  {\sigma^\mu}^{a{\bar a}},\nonumber \\ &{g^\mu}_{{\bar a}a} := ({\tilde \sigma}^\mu{)^T}_{{\bar a}a},~~~~\,g_{\mu{\bar a}a} := ({{\tilde \sigma}_\mu)^T}_{{\bar a}a}.
\end{align}
These symbols can be consistently manipulated according to the rules of both spinor and spacetime index gymnastics. In order to obtain the numerical values of the van der Waerden symbols, and when translating expressions into expressions involving matrix operations the identifications in Eq. (\ref{vdw}) must be used. Given this recipe for converting spinor indices to spacetime indices and vice versa, we obtain the following useful relations
\begin{align}
&g_{\mu\nu} = {g_\mu}^{a{\bar a}}{g_\nu}^{b{\bar b}}\omega_{ab}\omega_{{\bar a}{\bar b}}= g_{\mu a{\bar a}}{g_\nu}^{a{\bar a}},\nonumber \\ &g^{\mu\nu} = {g^\mu}_{{\bar a}a}{g^\nu}_{{\bar b}b}\omega^{ab}\omega^{{\bar a}{\bar b}}={g^\mu}_{{\bar a}a}{g^\nu}^{{\bar a}a},\nonumber \\ 
&g_{{\bar a}a{\bar b}b} := {g^\mu}_{{\bar a}a}g_{\mu{\bar b}b} =\omega_{ab}\omega_{{\bar a}{\bar b}},\nonumber \\ &g^{a{\bar a}b{\bar b}} :=g^{\mu a{\bar a}}{g_{\mu}}^{b{\bar b}} = \omega^{ab}\omega^{{\bar a}{\bar b}}
\end{align}
where the above expressions involving the symbols $g_{{\bar a}a{\bar b}b}$ and $g^{a{\bar a}b{\bar b}}$ can be viewed as definitions.

In general a spacetime index $\mu$ will correspond to a composite spinor pair of indices $a{\bar a}$ or ${\bar a}a$. It should therefore be evident how spacetime tensors can be constructed from spin-tensors. Regarding the representation theory of the Lorentz group ${\mathcal L}_+^\uparrow$, the carrier space of the $(i,i)$'th representation of $Sp(2,{\mathbb C})$ can be viewed as the space of totally symmetric tensors over spacetime with components $h^{(\mu_1...\mu_{2i})}$ or $h_{(\mu_1...\mu_{2i})}$. Given the above translation scheme it is clear that this is consistent with the specification given previously for the general $(i,j)$ representations of $Sp(2,{\mathbb C})$ in terms of spin-tensors.

\section{Proof that the generalised BC implies there is no current normal to the plates}\label{B}

We wish to prove that for an arbitrary spin-$\nicefrac{n}{2}$ field there exists a BC, which implies $n_\mu(0,d)j^\mu(0,d;n)=0$ where the local current $j^\mu(n)$ is defined in Eq. (\ref{currentm}). We consider the plate at $x^3=0$ for simplicity. The outward pointing normal to the surface $x^3=0$ has components $n^\mu=(0,0,0,-1)$. The generalised BC we choose (for $x^3=0$) is according to Eq. (\ref{gen})
\begin{align}\label{bc}
{\sigma^3}_{{\bar a}_1a_1}...{\sigma^3}_{{\bar a}_na_n}\psi^{a_1...a_n} = \psi_{{\bar a}_1...{\bar a}_n},
\end{align}
and we wish to show that this implies
\begin{align}
n_\mu(0)j^\mu(0) ={\sigma^3}_{{\bar a}_1a_1}{\sigma^0}_{{\bar a}_2a_2}...{\sigma^0}_{{\bar a}_na_n}\psi^{a_1...a_n}\psi^{{\bar a}_1...{\bar a}_n} = 0.
\end{align}
Now, Eq. (\ref{bc}) holds if and only if
\begin{align}\label{y}
\omega^{{\bar a}_1{\bar a}_1'}{\sigma^3}_{{\bar a}_1'a_1}...\omega^{{\bar a}_n{\bar a}_n'}{\sigma^3}_{{\bar a}_n'a_n}\psi^{a_1...a_n} = \psi^{{\bar a}_1...{\bar a}_n}.
\end{align}
Substituting in place of the factor $\psi^{{\bar a}_1...{\bar a}_n}$ contained in $j^\mu(0)$ the left-hand-side of Eq. (\ref{y}), we obtain
\begin{align}\label{z}
&n_\mu(0)j^\mu(0)\nonumber \\ &~~={\sigma^3}_{{\bar a}_1a_1}{\sigma^0}_{{\bar a}_2a_2}...{\sigma^0}_{{\bar a}_na_n} {(\omega\sigma^3)^{\bar a_1}}_{a_1'}...{(\omega\sigma^3)^{\bar a_n}}_{a_n'}\nonumber \\ &~~~~~~\times \psi^{a_1'...a_n'}\psi^{a_1...a_n}\nonumber \\ &~~= {(\sigma^3\omega \sigma^3)}_{a_1a_1'} (\omega\sigma^3)_{a_2 a_2'}...(\omega\sigma^3)_{a_na_n'}\psi^{a_1'...a_n'}\psi^{a_1...a_n}
\end{align}
where $\sigma^3 =(\sigma^3)^T$ and $\sigma^0 =(\sigma^0)^T=I$ have been used. Using $2\sigma^3\omega\sigma^3=-\omega$ and $\omega\sigma^3 =-\sigma^1$, Eq. (\ref{z}) gives
\begin{align}\label{x}
2n_\mu(0)j^\mu(0)&=\mp \omega_{a_1a_1'} {\sigma^1}_{a_2 a_2'}...{\sigma^1}_{a_n a_n'}\psi^{a_1'...a_n'}\psi^{a_1...a_n} \nonumber \\ &= \mp {\sigma^1}_{a_2 a_2'}...{\sigma^1}_{a_n a_n'}\psi^{a_1'...a_n'}{\psi_{a_1'}}^{a_2...a_n} \nonumber \\
&= \pm  {\sigma^1}_{a_2 a_2'}...{\sigma^1}_{a_n a_n'}{\psi_{a_1'}}^{a_2'...a_n'}\psi^{a_1'a_2...a_n}
\end{align}
where the $-$ sign on the top line corresponds to $n$-odd (fermions) and the $+$ sign to $n$-even (bosons). Relabeling the indices $a_i \leftrightarrow a_i',~i=2,...,n$ and using $\sigma^1 = (\sigma^1)^T$ the last line above gives
\begin{align}
2n_\mu(0)j^\mu(0)=\pm {\sigma^1}_{a_2 a_2'}...{\sigma^1}_{a_n a_n'}\psi^{a_1'...a_n'}{\psi_{a_1'}}^{a_2...a_n},
\end{align}
which is the negative of the second line in Eq. (\ref{x}). This proves that the generalised BC in Eq. (\ref{bc}) implies $n_\mu(0)j^\mu(0)=0$.








\end{appendix}

\end{document}